\documentclass[,aps,prx,twocolumn,floatfix,citeautoscript,showpacs,notitlepage,superscriptaddress]{revtex4-2}
\usepackage[english]{babel}
\usepackage[pdftex,colorlinks=false]{hyperref}
\addto\extrasenglish{%

}%
\addto\captionsenglish{}
\usepackage{latexsym}
\usepackage{amsfonts}
\usepackage{amsmath,amssymb,amsthm}
\usepackage{bbm}
\usepackage{mathtools}
\usepackage[dvipsnames]{xcolor}
\usepackage{graphicx}
\usepackage{bm}
\usepackage{multirow}
\usepackage{txfonts}
\usepackage{ulem}
\usepackage{lipsum}
\usepackage{blindtext}
\usepackage{subfiles}
\usepackage{amssymb}
\usepackage{amsmath}
\usepackage{eurosym}
\let\origunderline\underline
\renewcommand{\underline}[1]{\origunderline{\smash{#1}}}

\newcommand{\bs}[1]{{\boldsymbol{#1}}}

\renewcommand{\i}{\mathrm{i}}
\newcommand{\e}{\mathrm{e}}

\newcommand{\bk}{\bs{k}}
\newcommand{\bq}{\bs{q}}

\newcommand{\mf}{\overline}

\newcommand{\ket}[1]{\left| #1\right\rangle}
\renewcommand{\mathbb}{\mathbbm}

\newcommand{\app}[1]{Appendix~\ref{#1}}

\renewcommand{\dag}{\dagger}
\newcommand{\nodag}{{\vphantom\dag}}

\DeclareMathOperator{\Tr}{Tr}

\let\Im\relax

\DeclareMathOperator{\Im}{Im}

\definecolor{dr}{rgb}{0, 0, 1}
  
\usepackage{letltxmacro}
\LetLtxMacro{\originaleqref}{\eqref}
\renewcommand{\eqref}{Eq.~\originaleqref}

\begin{document}

%-------------------- Title, authors and abstract ---------------------
%%%%%%%%%%%%%%%%%%%%%%%%%%%%%%%%%%%%%%%%%%%%%%%%%%%%%%%%%%%%%%%%%%%%%%%
	
\title{Fluctuation corrections to the free energy of strongly correlated electron systems}

	\author{David Riegler}
	\email{riegler.physics@gmail.com}
	\affiliation{Institute for Theory of Condensed Matter, Karlsruhe Institute of Technology, D-76128 Karlsruhe, Germany}
        \affiliation{Institute for Theoretical Physics, University of W\"urzburg,
        D-97074 W\"urzburg, Germany}

    \author{Jannis Seufert}
    \affiliation{Institute for Theoretical Physics, University of W\"urzburg,
        D-97074 W\"urzburg, Germany}
    \affiliation{Würzburg-Dresden Cluster of Excellence ct.qmat, University of W\"urzburg,
        D-97074 W\"urzburg, Germany}

    \author{Ronny Thomale}
    \affiliation{Institute for Theoretical Physics, University of W\"urzburg,
        D-97074 W\"urzburg, Germany}
    \affiliation{Würzburg-Dresden Cluster of Excellence ct.qmat, University of W\"urzburg,
        D-97074 W\"urzburg, Germany}
  
	\author{Peter W\"olfle}
  	\email{peter.woelfle@kit.edu}
	\affiliation{Institute for Theory of Condensed Matter, Karlsruhe Institute of Technology, D-76128 Karlsruhe, Germany}
    \affiliation{Institute for Quantum Materials and Technologies, Karlsruhe Institute of Technology, D-76021 Karlsruhe, Germany}
 
 \date{\today }
	
%%%%%%%%%%%%%%%%%%%%%%%%%%%%%%%%%%%%%%%%%%%%%%%%%%
%%%%%%%%%%%%%%%%%%%%%%%%%%%%%%%%%%%%%%%%%%%%%%%%%%

\begin{abstract}
We determine the free energy of strongly correlated electron systems in the example of the Hubbard model by calculating the contribution of spin and charge fluctuations to the Gutzwiller approximation mean field result. We employ the slave boson formulation of Kotliar and Ruckenstein in its spin-rotation invariant form in the usual continuous time approximation of the functional integral  representation, corrected by ``high frequency contributions" (SRIKR+). Previous method-related shortcomings are shown to be overcome when the correct operator ordering for the renormalized kinetic energy is used. The results for the ground state energy in the paramagnetic phase are in very good agreement with state-of-the-art results obtained by methods such as density matrix embedding theory (DMET), quantum Monte Carlo (QMC) and others. The leading low temperature behavior of the free energy allows to extract the quasiparticle effective mass, in particular its enhancement near a continuous phase transition into an ordered state. Our work demonstrates that the SRIKR+ method is competitive with the best available alternative methods and equips the slave-boson approach with an improved synoptic power to explore strongly correlated electron systems.  
\end{abstract}
	
\maketitle
	
\section{Introduction}

%%---------------------------------------------------------------------
%%---------------------------------------------------------------------
%% INTRODUCTION
%%---------------------------------------------------------------------
%%---------------------------------------------------------------------

Strongly correlated electron systems in metals have been in the focus of condensed matter research for decades \cite{hubbard1963electron,hubbard1964,hubbard1965electron,Kanamori1963,Lee1986}. These systems show a remarkable variety of emergent phenomena such as ordered states in the charge and spin sector, superconductivity, and more, as well as correlated states without apparent order parameter \cite{anderson1984basic}. The theoretical description of these phenomena is beyond the reach of perturbation theory, although judicious resummations of perturbation theory such as provided by functional renormalization group methods may allow to get as far as to the intermediate interaction strength regime \cite{RevModPhys.84.299,Beyer2022}. A powerful tool to calculate a large class of properties of strongly correlated systems is dynamical mean field theory (DMFT) \cite{PhysRevLett.62.324,RevModPhys.68.13}, which maps the system to a quantum impurity system allowing for accurate numerical evaluation. Quantum Monte Carlo methods offer an unbiased principally exact access to thermodynamic properties, but are limited by the fermion sign problem and the system size problem \cite{gubernatis2016quantum}. Methods based on the density matrix renormalization group (DMRG) likewise are limited by accessible system sizes \cite{verstraete2023density}. More recently, Quantum Embedding (QE) theories have been devised, combining methods sufficiently powerful to deal with strong correlations and with details of the electronic and atomic structure of a solid \cite{KentKotliar2018,SunChan2016,Wouters2016}. 

On an elementary level, the problem of strong electronic correlations arises because the behavior of electrons in such systems depends on the local occupation state they are in: for example in a single electronic band situation whether the atomic level the electron resides in is singly or doubly occupied. Very early on, Gutzwiller introduced an ingenious non-perturbative approximation method using the statistical probabilities for the occupation of these four states to devise an approximate evaluation of the energy of a variational state \cite{gutzwiller1963}.  We recall that this so-called Gutzwiller approximation (GA) provides the exact evaluation of the variational free energy in the limit of infinite dimensions \cite{PhysRevLett.62.324} . This approach has been very successful in presenting a first qualitatively correct theory of the Mott metal-insulator transition \cite{PhysRevB.2.4302,RevModPhys.56.99}.
It also provided almost quantitative results for the thermodynamic properties of liquid Helium-3, a model system of a strongly correlated Fermi liquid \cite{PhysRevB.35.6703}. Recently an extension of the GA, denoted as the ghost Gutzwiller approximation (gGA) \cite{Dobrosav2017,Dobrosav2021} has been developed. The gGA framework incorporates auxiliary fermionic degrees of freedom to enrich the variational
space. gGA has been shown to be similar in accuracy to DMFT \cite{Dobrosav2017,Dobrosav2021,Fabrizio2023,LeeMelnick2023,LeeLanata2023}, 

In order to represent the information about the state of occupancy of an electron in a strongly correlated system in a more direct way, the concept of slave bosons has been invented (for reviews see \cite{Fresard2012,Fresard2015}) . In the simplest case, the theory needs to represent four states per lattice site: a doubly occupied state $\ket{2}$, two singly occupied states $\ket{\uparrow}$ and $\ket{\downarrow}$ (spin up or down), and the empty state $\ket{0}$. Early slave boson representations defined two bosons for the states $\ket{0}$ and $\ket{2}$ and two fermions for the states $\ket{\uparrow}$ and $\ket{\downarrow}$, or vice versa \cite{SEBarnes_1976,PhysRevB.29.3035}. The problem with this choice is that spin and charge degrees of freedom are not handled on equal footing. This deficiency has been removed by Kotliar and Ruckenstein (KR), who proposed to represent each of the four states by its own boson and in addition two pseudo-fermions to display the fermionic nature of the electron \cite{kotliar_new_1986}. The electron operator is thus defined by a product of the pseudo-fermion operator and a boson transition operator (encoding the information about the initial and final states out of the four possibilities). 

These improvements themselves would not yet lead to good results. The decisive ingredient making the KR-approach into a quantitatively competitive method is that the mean field solution may be dramatically improved by adding renormalization operators to the bosonic part of the electron operator designed such as to lead to the Gutzwiller mean field result \cite{kotliar_new_1986}. These operators reduce to unity in the physical subspace and thus do not violate the faithfulness of the representation. The initial KRSB-method did not respect the rotation symmetry in spin space. In other words, when considering fluctuations about the mean-field solution, only longitudinal spin fluctuations were accessible. This weakness was corrected by adding two more slave bosons to represent the singly occupied states as either a combination of a spin $0$ boson or a spin $1$ vector boson with the spin $\frac12$ pseudofermions to form a spin $\frac12$ electron \cite{woelfle_spin_rotation_1989,woelfle_spin_1992,Fresard2012}. In this way, the transverse spin fluctuations may be naturally included. A generalization of the model to $N$ orbitals has been introduced in \cite{woelfle_spin_1992}, where it is shown that the mean field solution becomes exact in the limit $N\rightarrow\infty$.

 In order to calculate correlation functions and the contribution of fluctuations to the free energy (or the grand potential), it is useful to map the quantum many-body expectation value of the statistical operator to a functional integral over coherent states of the boson and fermion operators involved. The ordering properties of the quantum operators are preserved by introducing a (imaginary) time $\tau_{n}=n\beta /M$, $n=1,...,M$ labeling time slices. Here, $\beta=1/T$ is the inverse of temperature $T$ (we use units in which Boltzmann's constant $k_{B}=1$) and $M>>1$ is a number which will be taken to infinity at the end of the calculation (limit of continuous time (CT)). Calculations in the discrete-time representation are cumbersome and are usually avoided in favor of taking the CT limit at the outset of the calculation. For a recent exact evaluation of the 2-site extended Hubbard model, see Ref.~\cite{dao2024exact}. The CT approximation is legitimate when calculating correlation functions \cite{JWRasul_1988,PhysRevB.41.142,Li1991,woelfle_spin_1997}, as we show below. As far as the calculation of the free energy is concerned, however, the evaluation in the CT limit leads to huge errors that need to be corrected \cite{PhysRevB.44.2403,arrigoni1994functional}. In the following we use a ``poor man's way'' of calculating these ``high frequency contributions'' (HFC), termed so because these arise from contributions at high frequency. This method was introduced by Arrigoni and Strinati \cite{PhysRevB.52.2428} and was checked against the full calculation in the discrete-time formulation. The corrections appear because in the CT limit the ordering of non-commuting operators is lost. We find that applying the HFC terms reduces the magnitude of the fluctuation corrections to a small fraction of the initial CT result, to values in excellent agreement with benchmark results from other state-of-the-art methods.

 In this paper, we calculate fluctuation corrections to the free energy of the Hubbard model, to be concrete. The method of calculation mapped out in the following may be applied to any other model of correlated fermions, though. In contrast to the ghost Gutzwiller approach we do not improve the mean-field solution by expanding the variational space, but we calculate the contribution of fluctuations about the initial GA mean field. 
 
 We begin by summarizing the spin-rotation-invariant Kotliar-Ruckenstein slave-boson representation (SRIKR). This includes the functional integral representation in discrete-time form subjected to gauge transformations removing the phases of five of the six complex-valued boson fields. As a first step, the mean-field (saddle point) solution is found, which in the paramagnetic state may be done in large part analytically. Next, the action is expanded up to second order in the fluctuating fields about the saddle point solution. Symmetry dictates that the $10\times 10$ fluctuation matrix decomposes into a $4\times 4$ block in charge space and three identical $2\times 2$ blocks in spin space. 

 The fluctuation contribution to the partition function may now be evaluated in the CT approximation, separately for each of the subsectors (charge, spin). Special attention is paid to the treatment of the radial variables. It is shown that in both the spin and charge sectors the result of the integration is equivalent to the usual procedure assuming real fields (up to irrelevant prefactors). In the limit of the vanishing interaction $U=0$ the contribution of fluctuations is expected to vanish, which is recovered in the spin sector, but not in the charge sector. This is a first serious indication that something is missing in the CT limit. On top of that, the magnitude of the fluctuation contribution is found to be two orders of magnitude too large. 

 To cure this deficiency of the continuous-time treatment of the functional integral, we calculate contributions as mentioned above, such that its evaluation becomes exact. These HFC terms are shown to restore the correct $U=0$ limit. Moreover, the HFC terms reduce the total contribution of fluctuations to the free energy to the order of a few percent or less of the mean-field value. We shall call this method SRIKR+ in the following. A comparison with high-precision results obtained via density matrix embedding theory (DMET) and other methods demonstrates very good agreement, considering the uncertainties of the DMET regarding different cluster sizes and shapes. The SRIKR+ data points for the free energy are found to be systematically lower than the above-mentioned benchmark results. Here the ``density corrections" necessary when calculating the free energy out of the grand potential have been applied. We explore probable reasons for the discrepancy of the SRIKR+ results (valid in the thermodynamic limit) and the DMET results for small impurity clusters.

 In addition to the ground-state energy we calculate the low-temperature behavior of the free energy. The leading $T^2$ contribution allows us to determine the limiting low-temperature value of the specific heat. In the framework of Fermi liquid theory, the specific heat, in turn, reflects the quasiparticle effective mass ratio $m^\ast/m$. Within the present slave boson formulation, a first estimate of $m^\ast/m$ is obtained in mean-field approximation. As we show below, the contribution of fluctuations may give rise to substantial corrections of the mean field effective mass value, in particular near a continuous phase transition where fluctuations get large. 
 
\section{Spin-rotation invariant Kotliar-Ruckenstein slave-boson representation}

%%---------------------------------------------------------------------
%% Model: Hamiltonian of the extended Hubbard model
%%---------------------------------------------------------------------

 We employ the one-band $t$\,--\,$t'$\,--\,$U$ Hubbard model on the 2D square lattice with on-site repulsion $U$,
 where the operator $c_{i,\sigma }^{\dagger }$ creates an electron with spin $\sigma =\{\uparrow ,\downarrow \} $ at site $i$, and $n_i= \sum_\sigma c_{i,\sigma }^{\dagger}c^\nodag_{i,\sigma}$ is the density operator
 	\begingroup
	\allowdisplaybreaks
	\begin{align}\label{eq:Hamiltonian}
		\begin{split}
			\hat{H}= &- \sum_{\sigma=\uparrow,\downarrow}
			\left(
			t^\nodag_{\vphantom i} \sum_{\langle i,j \rangle_1}  \, c_{i,\sigma }^{\dagger}c^\nodag_{j,\sigma}  
			+ t{\vphantom i}'  \sum_{\langle i,j \rangle_2} c_{i,\sigma }^{\dagger}c^\nodag_{j,\sigma}
			+\mathrm{h.c.}
			\right) \\
			&- \mu_0 \sum_i n_{i}  
			+U\sum_{i} c_{i,\uparrow }^\dagger c_{i,\uparrow}^\nodag c_{i,\downarrow }^\dagger c_{i,\downarrow }^\nodag .
		\end{split}
	\end{align}
	\endgroup
	 Moreover, $\langle i,j \rangle_n$ denotes an $n$th nearest-neighbor pair, and $\mu_0$ is the chemical potential. 
	
%---------------------------------------------------------------------
% Slave-boson representation
%---------------------------------------------------------------------

We apply the spin-rotation-invariant Kotliar-Ruckenstein slave-boson (SRIKR-SB) representation ~\cite{kotliar_new_1986,woelfle_spin_1992}, whereby we introduce the bosonic fields $d_{i},e_{i},p_{0,i}$ and $\mathbf{p}_{i}=(p_{1,i},p_{2,i},p_{3,i})$, which create/annihilate doubly, empty, and singly occupied states, respectively, as well as a set of auxiliary fermionic fields $\bs f_{i}^{\nodag}=(f_{i,\uparrow}^{\nodag},f_{i,\downarrow}^{\nodag})$ and Lagrange multiplier fields $\alpha_i, \beta_{0,i}, \bs \beta_{\mu,i}$, $\mu=1,2,3$ to recover the physical subspace via constraints. 
The Hamiltonian, including the Lagrange multiplier terms allowing to enforce the constraints, reads 
\begin{align}\label{eq:KRHam}
%\begin{split}
\hat{H} & =\sum_{i,j,\sigma ,\sigma ^{\prime}}f_{i,\sigma }^{\dag }
\left\{t_{i,j}\underline{z}_{i}^{\dag }\underline{z}_{j}-\delta _{ij}\left[(\mu _{0}-\beta _{0,i})\underline{\tau }_{0}-\bs\beta_{i}\cdot \underline{\bs\tau }\right]\right\}_{\sigma \sigma ^{\prime}}f_{j,\sigma ^{\prime }} \nonumber \\
& +\sum_{i}\Bigl\{-\beta _{0,i}\left(p_{0,i}^{\dag }p^\nodag_{0,i}+\mathbf{p}_{i}^{\dag }\cdot \mathbf{p}_{i}^\nodag+2d_{i}^{\dag }d_{i}^\nodag\right)-\bs\beta_{i}\cdot \left(p_{0,i}^{\dag }\mathbf{p}^\nodag_{i}+\mathbf{p}_{i}^{\dag }p^\nodag_{0,i}\right) \nonumber\\
& +\alpha _{i}\left(e_{i}^{\dag }e^\nodag_{i}+p_{0,i}^{\dag }p^\nodag_{0,i}+\mathbf{p}_{i}^{\dag }\cdot \mathbf{p}^\nodag_{i}+d_{i}^{\dag }d^\nodag_{i}-1\right)\Bigl\}.  
%\end{split}
\end{align}
The $z$-matrix factors are adopted in the operator ordering form defined originally by KR as 
\begin{align}\label{eq:KRzfactor}
\underline{z}& =\sqrt{2}\underline{L}\left[e^{\dag }\underline{p}+\widetilde{%
\underline{p}}^{\dag }d\right]\underline{R} \\
\underline{L}& =\frac{1}{\sqrt{(1-d^{\dag }d)\underline{\tau }_{0}-2%
\underline{p}^{\dag }\underline{p}}}, \\
\underline{R}& =\frac{1}{\sqrt{(1-e^{\dag }e)\underline{\tau }_{0}-2%
\widetilde{\underline{p}}^{\dag }\widetilde{\underline{p}}}},
\end{align}%
and the matrix operator $\underline{p}$ is defined in terms of $p_0$ and $\mathbf{p}=(p_{1},p_{2},p_{3})$\ as 
\begin{equation}
\underline{p}=\frac{1}{2}\left(p_{0}\underline{\tau }_{0}+\mathbf{p}\underline{%
\mathbf{\tau }}\right),
\end{equation}
where $\underline{\mathbf{\tau }}=(\tau _{1},\tau _{2},\tau _{3})$ is the vector of Pauli matrices and $\tau_{0}$ is the identity matrix. Here, $\widetilde{\underline{p}}$ is the time-reversed operator, for which $\mathbf{p}$ reverses sign. 

We note that the order of operators in the above definition of $\underline{z}$ is essential for the correct calculation of the high-frequency contributions below. This particular order is motivated by applying the weighting factors $\underline{L}$ and $\underline{R}$ to the initial and final states of the boson transition operators $e^{\dag }\underline{p}+\widetilde{\underline{p}}^{\dag }d$ constituting the bosonic part of the electron annihilation (or creation) operators.

The spin density related slave boson operators obey the following commutation relations 
\begin{eqnarray}
\left[ p_{\alpha \beta },p_{\gamma \delta }^{\dag }\right] &=&\frac{1}{2}\delta
_{\alpha\delta}\delta _{\beta\gamma}, \\
\left[ p_{\mu },p_{\nu }^{\dag }\right] &=&\delta _{\mu\nu}.
\end{eqnarray}%
We shall use 
\begin{eqnarray}
2\underline{p}^{\dag }\underline{p} &=&\frac{1}{2}\left[\left(p_{0}^{\dag }p_{0}+\mathbf{p}^{\dag }\mathbf{p}\right)\underline{\tau }_{0}+\left(\mathbf{p}^{\dag }p_{0}+p_{0}^{\dag }\mathbf{p}\right)\underline{\mathbf{\tau }}\right],  \\
2\widetilde{\underline{p}}^{\dag }\widetilde{\underline{p}} &=&\frac{1}{2}\left[\left(p_{0}^{\dag }p_{0}+\mathbf{p}^{\dag }\mathbf{p}\right)\underline{\tau }_{0}-\left(\mathbf{p}^{\dag }p_{0}+p_{0}^{\dag }\mathbf{p}\right)\underline{\mathbf{\tau }}\right],
\end{eqnarray}
in order to expand the $\underline{z}$-operator in terms of fluctuations about the mean field solution below. 

In order to calculate the thermodynamic properties of the system we first need to calculate the grand canonical partition function
\begin{equation}
Z=\Tr\exp\left[-\beta \left(\hat{H}-\mu_{0}\hat{N}\right)\right],
\end{equation}
where $\hat{H}$ is the Hamiltonian of the system, $\mu_{0}$ is the chemical
potential and $\hat{N}$ is the particle number operator and $\beta$ is the inverse temperature. In the present case, the expression for $Z$ has to be supplemented by projection operators that enforce the constraints on the slave-boson occupation numbers. The grand potential per lattice site is given by
\begin{equation}
\Omega=-\frac{T}{N_{L}}\log Z,
\end{equation}
where $N_{L}$ is the number of lattice sites.

\section{Static mean-field approximation}

The basic starting point of the calculation of $\Omega$ is the mean field (MF) approximation defined by replacing the slave boson operators and the constraint fields by site-independent real-valued numbers $d_{i}\rightarrow \overline{d}$, $e_{i}\rightarrow \overline{e}$, $p_{\mu,i}\rightarrow \overline{p}_{\mu}$, and the Lagrange multipliers by $\alpha_{i}\rightarrow \overline{\alpha}$, $\beta_{\mu,i}\rightarrow \overline{\beta}_{\mu}$, $\mu= 0,1,2,3$. Within this ansatz, the grand potential per lattice site is found as
\begin{equation}\label{eq:OmegaMF}
    \Omega^{(0)} = -\frac{2T}{N_{L}}\sum_{\bk}\ln\left[1+\e^{-\epsilon_\bk/T}\right] + U \overline{d}^2 -\beta_0 n^{(0)},
\end{equation}
where
\begin{align}
\epsilon_\bk &=  \overline{z}^2 \xi_{\bk} + \beta_0 -\mu_0,\\
\xi_\bk &= -2t\left(\cos k_x + \cos k_y\right) - 4t' \cos k_x \cos k_y,
\end{align}
and
\begin{equation}
n^{(0)} = - \frac{\partial \Omega^{(0)}}{\partial \mu_0}
\end{equation}
is the MF value of the density per site. The MF free energy is given by
\begin{equation}\label{eq:F0}
    F^{(0)} =\Omega^{(0)} + \mu_0 n^{(0)}.
\end{equation}
The mean field values are determined by finding the minimum of $F^{(0)}$ with respect to variations of the slave boson mean field values and simultaneously the maximum w.r.t.~variations of the Lagrange parameters. In the paramagnetic state at density $n$ (mean electron occupation per site) or equivalently at hole doping $\delta$ in the half filled system, a semi-analytical solution is available at any interaction strength $U$ (see \app{app:sec:MF}) and the independent number of MF variables can be reduced with the introduction of an effective chemical potential \cite{Hubbard_Wuerzburg}. 

%---------------------------------------------------------------------
% MF approximation
%---------------------------------------------------------------------

\section{Fluctuations around the saddle point}

We calculate Gaussian fluctuations around the MF saddle point by expanding the bosonic fields up to the second order. In order to do so, we apply the functional integral representation of the partition function. As we highlight in \autoref{sec:discrete}, this formalism is only exact in the discrete imaginary time formulation, where the limit to continuous imaginary time (CT) has to be taken at the very end of the calculation. Taking the CT limit in advance greatly simplifies the calculation but leads to huge errors for the fluctuation corrections of the grand potential $\Omega^{(2)}$ \cite{PhysRevB.44.2403,arrigoni1994functional}. 
The charge and spin susceptibility $\chi_c(q)$ and $\chi_s(q)$, however, are unaffected by the HFC such that the results of previous works \cite{JWRasul_1988,PhysRevB.41.142,Li1991,woelfle_spin_1997} are still correct. We calculate the CT and HFC contributions in \autoref{sec:CT} and \autoref{sec:HFC}, respectively, and prove the validity of this approach with the exactly solvable test case of free bosons in \app{app:sec:freebosons}. In summary, the grand potential up to second order is then given by
\begin{equation}
    \Omega = \Omega^{(0)}+\Omega^{(2)} = \Omega^{(0)}+ \Omega^{(2)}_{CT} + \Omega^{(2)}_{HFC},
\end{equation}
where the MF contribution $\Omega^{(0)}$ is given by \eqref{eq:OmegaMF} and the CT and HFC fluctuation contributions are given by \eqref{eq:OmegaCTa} and \eqref{eq:OmegaHFC}, respectively. The free energy up to second order is recovered with the Legendre transformation
\begin{align}\label{eq:F2}
    F &= \Omega - \mu_0 \frac{\partial \Omega }{\partial \mu_0}= \Omega^{(0)}+\Omega^{(2)} + \mu_0 \left(n^{(0)} + n^{(2)}\right)  
\end{align}
where $n^{(0)}$ is the MF density and the fluctuation correction to the density is given by
\begin{equation}\label{eq:n2}
n^{(2)}  = - \frac{\partial \Omega^{(2)}}{\partial \mu_0}.
\end{equation}

\subsection{Functional integral representation of the partition function}\label{sec:discrete}

The operator-valued expression of the partition function cited above may be transformed into a functional integral
over coherent state amplitudes of the form of complex-valued numbers (bosons) and Grassmann fields (fermions).
It is necessary to indicate the ordering of
the various terms, as required in the operator language. For that reason, the fields are endowed with a ``time" label $\tau_{m}$, representing the position of the field in the interval $[0,\beta]$, with $\beta=1/T$. This interval is subdivided into $M$ slices, $m=1,2,...,M$. Eventually, the limit $M\rightarrow\infty$ is taken,
the so-called continuum time limit.  We first consider the functional integral in discrete-time representation. The general form of the partition function
involving field amplitudes $\xi_{\alpha},\xi_{\alpha}^{\ast}$ representing the annihilation and creation operators in the problem is given by (see, e.g., Ref.~\cite{negele1988quantum})
\begin{equation}\label{eq:Zdiscrete}
Z=\lim_{M\rightarrow\infty}\prod_{m=1}^{M}\prod_{\alpha}\mathit{D[\xi
_{\alpha,m},\xi_{\alpha,m}^{\ast}]}e^{-S[\xi_{\alpha,m},\xi_{\alpha,m}^{\ast
}]},
\end{equation}
where the action is expressed as
\begin{align}\label{eq:Sdiscrete}
\begin{split}
S\left[\xi_{\alpha,m}^*, \xi_{\alpha,m} \right]
= &\eta \sum_{m=2}^M\sum_\alpha \left[\xi^*_{\alpha,k}\frac{\xi_{\alpha,m}-\xi_{\alpha,m-1}}{\eta}+H\left(\xi^*_{\alpha,m},\xi^\nodag_{\alpha,m-1} \right)\right] \\ 
+&\eta \sum_\alpha\Bigg[ \xi^*_{\alpha,1} \frac{\xi_{\alpha,1}-\zeta\xi_{\alpha,M}}{\eta} +H\left(\xi^*_{\alpha,1},\zeta\xi^\nodag_{\alpha, M} \right)  \Bigg],
\end{split}
\end{align}
where $\zeta=1$ for bosons and $\zeta=-1$ for fermions and $\eta=\beta/M$ is
the size of the time slice. The limit $M \rightarrow \infty$ should be taken at the end of the calculation. 
Calculations in the discrete-time representation are cumbersome and have been rarely done. Usually, the limit $M \rightarrow \infty$ is taken right at the beginning, leading to the familiar continuous-time representation considered next. As we shall see, the error incurred by this approximation in the calculation of $\Omega$ is unacceptably high and must be corrected by ``high frequency contributions (HFC)" as will be shown below.

\subsection{Continuous time (CT) functional
integral representation of the partition function}\label{sec:CT}

In the limit of continuous time, the action adopts the more familiar form 
\begin{align}
\begin{split}
S\left[\Phi^{\ast},\Phi\right]  &  =\int_{0}^{\beta}d\tau L\left[\Phi^{\ast}(\tau),\Phi
(\tau)\right],\\
L[\Phi^{\ast}(\tau),\Phi(\tau)]  &  =\sum_{j}\left\{\Phi_{j}^{\ast}(\tau
)\partial_{\tau}\Phi_{j}(\tau)+H[\Phi_{j}^{\ast}(\tau),\Phi_{j}(\tau
)]\right\},
\end{split}
\end{align}
where $j$ denotes lattice sites and $\Phi(\tau)$ is a vector of field amplitudes. The vector of fields $\Phi$ consists of the slave boson fields $e,d,p_{0},p_{1},p_{2},p_{3}$, the constraint fields $\alpha,\beta_{0},\beta_{1},\beta_{2},\beta_{3}$, and the pseudo-fermion fields $f_{\sigma}$. The slave boson fields are complex fields, to begin with, while the constraint fields are real valued. We may gauge away the phases of all boson fields, except for one, which we choose to be $d=d_{1}+id_{2}$, where $d_{1},d_{2}$ are real valued. The remaining boson fields are radial fields, meaning that they are real and positive fields and
the integration measure is, e.g., $d p_0^2 =2p_0 dp_0$ . The significance of radial fields in this context has been elucidated in Refs.~\cite{fresard2001slave,fresard2007slave}. 
In later evaluations we will eliminate the field $e$ by using the constraint
$e^{2}=1-d_{1}^{2}-d_{2}^{2}-p_{0}^{2}-\mathbf{p}^{2}$, where $\mathbf{p}%
^{2}=p_{1}^{2}+p_{2}^{2}+p_{3}^{2}$.
The Lagrangian has a bosonic and a fermionic part
\begin{equation}
L[\Phi^{\ast}(\tau),\Phi(\tau)]=L_{B}+L_{F}.
\end{equation}
The bosonic part reads 
\begin{align}
\begin{split}
L_{B}  &  =\sum_{j}\Big{[} U\left(d_{1,j}^{2}+d_{2,j}^{2}\right)+id_{1,j}\partial_{\tau}d_{2,j}-id_{2,j}\partial_{\tau}d_{1,j}\\
&  -\beta_{0,j}\left(p_{0,j}^{2}+\mathbf{p}_{j}^{2}+2d_{1,j}^{2}+2d_{2,j}%
^{2}\right)-2p_{0,j}\bs\beta_{j}\cdot\mathbf{p}_{j}\Big{]}.
\end{split}
\end{align}
Here we have already eliminated $e$ and $\alpha$. It has been shown that this elimination does not change the result of the CT contributions but considerably simplifies the calculation \cite{Hubbard_Wuerzburg}.
The fermionic part is obtained after integrating out the pseudo-fermion fields
\begin{align}
\begin{split}
L_{F}  =-\Tr\ln\Big{[}&\left(\partial_{\tau}-\mu_{0}%
+\beta_{0,i}\right)\delta_{\alpha\beta}\delta_{ij}+\bs\beta_{i}\cdot\bs\tau_{\alpha\beta}\delta_{ij}\\
&+t_{i,j}\sum_{\gamma}z_{i,\alpha\gamma}^{\dag}z^\nodag_{j,\gamma\beta}\Big{]},
\end{split}
\end{align}
where trace runs over the degrees of freedom of the pseudo-fermions. 

Within the scope of Gaussian fluctuations, we expand the action up to the second order in bosonic field fluctuations $\delta\psi_{\nu,q}$ around the MF saddle point with the MF action $\mathcal{S}^{(0)}$. It should be mentioned that calculation of Gaussian fluctuation contributions amounts to a one-loop order in the $1/N$ expansion of an extension of the Hubbard model to $N$ orbitals \cite{woelfle_spin_1992}. The action is then defined as
\begin{equation}
 \mathcal{S} = \mathcal{S}^{(0)} + \frac{T}{N_{L}}\sum_{\i\omega_n}\sum_{\bq} \delta\psi_{\mu,-q} \mathcal{M}_{\mu\nu}(q) \delta\psi_{\nu,q} +...    
\end{equation}
The fluctuation matrix is given by
\begin{equation}
\mathcal{M}_{\mu\nu}(q) = \frac12 \frac{\delta ^2 \mathcal{S}}{\delta \psi_{\mu,-q} \delta \psi_{\nu,q}},
\end{equation}
where $q=(\i\omega_n,\bq)$, with $\omega_n = 2\pi n T$ being a bosonic Matsubara frequency and $\bq$ the momentum vector of the fluctuations. In the CT limit, the Matsubara summation is given by 
$\sum_{\i\omega_n}=\sum_{n=-\infty}^{\infty}$. The fluctuation basis, $\delta\psi_{\nu,q} = \psi_{\nu,q} - \mf\psi_\nu$ denotes the deviation from the MF value $\mf\psi_\nu$ and the index $\nu$ labels the fluctuation fields. The fluctuation matrix is block-diagonal, whereby the charge sector with the basis $\delta \bs \psi_c = (d_1,d_2,p_0,\beta_0)^\top$ and the longitudinal/transversal spin sectors $\delta \bs \psi_{s,\mu} = (p_{\mu},\beta_{\mu})^\top$, with $\mu=1,2,3$ are decoupled in the paramagnetic state, where the three spin sectors are identical. In position-time space, the real-valued variables $d_1,d_2,\beta_0,\beta_\mu$ ($\mu=1,2,3$) are integrated from $-\infty \rightarrow \infty$, whereas the radial variables $p_0,p_\mu$ are integrated from  $0 \rightarrow \infty$ with integration measure $p_0dp_0$, etc. After taking the Fourier transform, however, all boson amplitudes and constraint fields become complex-valued and their real and imaginary parts are integrated from $-\infty \rightarrow \infty$. This is discussed in  \app{app:sec:functionslintegrals}. Therefore, the result of the integration of the fluctuation variables at a given $q$ is found as the inverse square-root of the determinant of the corresponding fluctuation matrix. In total, the determinant of the fluctuation matrix is given by
\begin{equation}
    \det\mathcal{M}(q) = \det\mathcal{M}_c(q) \left[\det\mathcal{M}_s(q)\right]^3.
\end{equation}
For the spin sector $\det\mathcal{M}_s\propto\chi_0/\chi_s$ holds, where the spin susceptibility satisfies $\chi_s\rightarrow\chi_0$ for $U\rightarrow 0$. Consequently the CT contributions are effectively determined by the spin susceptibility and vanish in the non-interacting limit. Although the non-interacting limit of the charge susceptibility is in full analogy: $\chi_c\rightarrow\chi_0$ for $U\rightarrow 0$, the charge determinant is not proportional to $\chi_0/\chi_c$. This may be traced to additional degrees of freedom in the charge sector, such as oscillations between densities of empty, singly, and doubly occupied sites that leave the average density $n=2d^2+p_0^2$ unchanged. We find a finite CT contribution at $U=0$ which, however, cancels out with the HFC.
The explicit expressions of $\mathcal{M}_s(q)$ and $\mathcal{M}_c(q)$ are provided in \app{app:sec:fluctuationmatrix} and the non-interacting limit is derived in \app{sec:app:U0}. 

By integrating out the fluctuation fields, we show in \app{app:sec:functionslintegrals} that the fluctuation corrections to the partition function are given by
\begin{equation}
    Z^{(2)}_{CT} = \prod_{q}\frac{1}{\sqrt{\det{\mathcal{M}(q)}}}.
\end{equation}
The contribution to the grand potential separates into a spin and charge contribution
\begingroup
\allowdisplaybreaks
\begin{align}
\Omega^{(2)}_{CT} &= \Omega^{(2)}_{CT,s} + \Omega^{(2)}_{CT,c},\label{eq:OmegaCTa}\\ 
    \Omega^{(2)}_{CT,s/c} &= \frac{a_{s/c}}{2}\frac{T}{N_{L}}\sum_{\bq}\sum_{\i\omega_n}\ln[\det\mathcal{M}_{s/c}(q)]\label{eq:OmegaCTs}, 
\end{align}
\endgroup
where $a_{s}=3$ and $a_{c}=1$. The numerical evaluation of the Matsubara summation requires care. Three different methods are discussed in \app{app:numerics}. 
We may map the Matsubara summation into a contour integral in the complex frequency plane, using the Bose distribution $n_B(\omega) = 1/[\exp(\omega/T)-1]$ as weight function, where the contour encircles the points $i \omega_n$ on the imaginary axis. The functions $\ln\det\mathcal{M}_{s/c}$ are analytic on the complex frequency plane except for the real axis. Next, the integration contour is moved to infinity and thereby wrapped around the real axis. The CT fluctuation contribution to the partition function is thus determined by integrating along the phase of the fluctuation determinant. We may exploit that the integrand is an even function in $\omega$ to reduce the integration to the positive frequency domain, yielding
\begin{equation}\label{eq:OmegaCT}
\Omega^{(2)}_{CT,s/c} = a_{s/c}\frac{1}{N_{L}}\sum_{\bq} \int_{0}^\infty \frac{d\omega}{2\pi} \coth \left(\frac{\omega}{2T}\right)\arg\left(\det\mathcal{M}_{s/c}\right).
\end{equation}

\subsection{High frequency contribution (HFC) to the continuous time functional integral representation of $Z$}\label{sec:HFC}

The approximation of continuous time in the functional integral for the partition function may be understood as doing a
calculation for a Hamiltonian for which the order of two non-commuting
operators is irrelevant. In other words the product of two Bose operators $%
b^{\dag }b$ is treated as if it were replaced by $(b^{\dag }b+bb^{\dag })/2$
. The difference is given by the commutator $b^{\dag }b=(b^{\dag }b+bb^{\dag
})/2+[b^{\dag },b]/2$. Therefore, in cases where the Hamiltonian is
effectively given by a bilinear operator expression, like in the case of
slave-boson representation in Gaussian fluctuation approximation, the
corrections to the free energy may be derived from the commutator terms in
the above expansion. Starting from the Hamiltonian in slave boson representation \eqref{eq:KRHam}, we may distinguish two types of contributions, of bosonic or  origin. The bosonic contribution may be directly read off the Hamiltonian, by replacing operator products $d_{i}^{\dag }d_{i}\rightarrow -1/2$ and so forth. The pseudo-fermionic part arises by expansion of the $z$-operators in terms of fluctuation operators $\delta d_{i} = d_{i}-\overline{d}$, and so forth, up to second order (Gaussian fluctuations). Higher-order fluctuations will be omitted in the spirit of a $1/N$ expansion in the limit $N\rightarrow \infty$.

\subsubsection{HFC in the spin sector}
The bosonic part of the HFC to the grand potential per site counting the three
spin orientations is given by 
\begin{equation}
\Omega_{HFC,3s,b}^{(2)}=3(\beta _{0}-\alpha )/2
\end{equation}%
The contribution to the grand potential from the pseudo-fermionic sector is found as 
\begin{equation}\label{eq:OmegaHFCs}
\Omega_{HFC,3s,f}^{(2)}=-2z_{HFC,3s,f}\overline{z}\varepsilon_0,
\end{equation}
where the characteristic pseudo-fermionic energy $\varepsilon_0$ is defined as
\begin{equation}\label{eq:epsilon0}
\varepsilon_0 = -\frac{2}{N_{L}}\sum_{\bk} n_F(\epsilon_\bk)\xi_\bk. 
\end{equation}
Here $n_F(\epsilon)=1/(\exp(\epsilon/T)+1)$ is the Fermi function and $\overline{z}$ is the mean field value of $z$.
The correction to the $z$-factor is found as
\begin{equation}\label{eq:ZHFCs}
z_{HFC,3s,f}=-\frac{3\overline{z}}{4}\left[\frac{1}{1-\overline{d}^{2}-\overline{p}_{0}^{2}/2}+\frac{1}{1-\overline{e}^{2}-\overline{p}_{0}^{2}/2}\right],
\end{equation}
for details, see \app{app:sec:HFC}.

\subsubsection{HFC in the charge sector}

The bosonic part of the HFC to the grand potential per site counting the three bosons $d,e,p_{0}$ is given by 
\begin{equation}
\Omega_{HFC,c,b}^{(2)}=-U/2+3(\beta _{0}-\alpha )/2.
\end{equation}%
From the pseudo-fermionic sector one finds 
\begin{equation}\label{eq:OmegaHFCc}
\Omega_{HFC,c,f}^{(2)}=-2z_{HFC,c,f}\overline{z}\varepsilon_{0}.
\end{equation}
The correction to the $z$-factor is given by
\begin{eqnarray}\label{eq:ZHFCc}
z_{HFC,c,f} &=&-\frac{\overline{z}}{2}\left[\frac{1}{1-\overline{d}^{2}-\overline{p}_{0}^{2}/2}+%
\frac{1}{1-\overline{e}^{2}-\overline{p}_{0}^{2}/2}\right], 
\end{eqnarray}
for details of the derivation see \app{app:sec:HFC}.

\subsubsection{Total result}

In total, the HFC to the grand potential may be summed up to be 
\begin{equation}\label{eq:OmegaHFC}
\Omega_{HFC}^{(2)}=-U/2+3(\beta _{0}-\alpha )+\Omega_{HFC,3s,f}+\Omega_{HFC,c,f}.
\end{equation}
The correction turns out to be substantial, of the order of the mean field result $\Omega_{0}$.
This is demonstrated in \autoref{fig:OmegaCTHFCs} and \autoref{fig:OmegaCTHFCc} showing the results of the CT and HFC evaluations separately.

%%---------------------------------------------------------------------
%%---------------------------------------------------------------------
%% RESULTS AT T=0
%%---------------------------------------------------------------------
%%---------------------------------------------------------------------

\section{Results at zero temperature}

\begin{figure}
    \centering
    \includegraphics[width=1\linewidth]{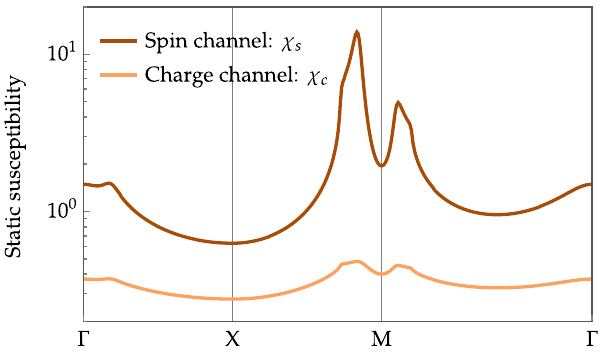}
    \caption{Static spin and charge susceptibility $\chi_{s/c}(\bs q, \omega =0)$ at $n=0.8$ for $U=2,t'=-0.2$ along the high symmetry lines of the Brillouin zone (BZ), i.e., $\Gamma =(0,0),\mathrm{X}=(\pi,0),\mathrm{M}=(\pi,\pi)$.}
    \label{fig:susceptability}
\end{figure}

In the present study, we investigate fluctuation corrections to paramagnetic ground states of the Hubbard model with next-nearest neighbor hopping $t'=0$ and $t'=\pm 0.2$. For small to moderate doping $\delta \lesssim 0.2$, we are therefore restricted to intermediate $U\lesssim 4$, to avoid magnetic instabilities. Full magnetic MF phase diagrams can be found in Refs.~\cite{fresard1992spiral,PhysRevB.44.7455,Fresard_1991,igoshev2013,igoshev2015,Seufert_2021,PhysRevB.110.085104}. The spin susceptibility $\chi_s$ diverges at the magnetic transition and is considerably enhanced in its vicinity, resulting in a significant fluctuation correction to the free energy. In contrast, the charge susceptibility is only slightly enhanced in the PM domain, since charge instabilities only appear at higher interaction strengths within the magnetic domain or with the addition of long-range interactions \cite{Seufert_2021,Riegler_2023}. This is exemplarily demonstrated in \autoref{fig:susceptability}, while comprehensive discussions of susceptibilities can be found in Refs.~\cite{JWRasul_1988,PhysRevB.41.142,Li1991,woelfle_spin_1997,PhysRevB.95.165127,Hubbard_Wuerzburg, Seufert_2021,Riegler_2023}. As a consequence, we consistently find higher fluctuation corrections from the spin channel compared to the charge sector, yielding a reduction of the ground-state energy compared to the MF value. The fluctuation corrections $F^{(2)}$ are determined according to \eqref{eq:F2} in addition to the KRSB MF value $F^{(0)}$, where we label the combined result $F^{(0)}+F^{(2)}$ as SRIKR+. Moreover, we consider density corrections according to \eqref{eq:n2}, which lead to a small correction of the magnetic MF phase boundaries in the $U-n$ plane. A generalization of SRIKR+ to non-local interactions is provided in \app{app:sec:V}.

\begin{figure*}
    \centering
    \includegraphics[width=0.99\linewidth]{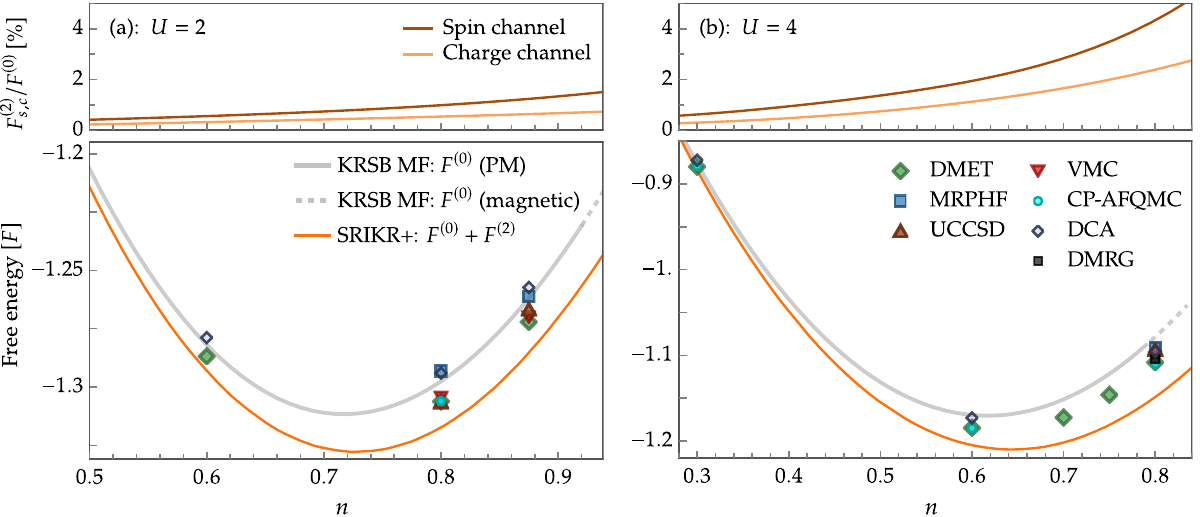}
    \caption{SRIKR+ free energy of the Hubbard model as a function of the density $n$ for $t'=0$ and the interactions (a): $U=2$, (b): $U=4$ along with benchmark data from different methods \cite{PhysRevX.5.041041}. The top panel displays the fluctuation corrections relative to the MF result, which consistently lower the ground state energy and increase towards half-filling and for higher interactions.}
        \label{fig:FEt20}
\end{figure*}

Since the exact ground state of the Hubbard model is unknown in 2D, approximation techniques at the cost of potential systematic errors are necessary. Therefore, we quantify our results by comparing the ground state energy with other state-of-the-art methods for strongly correlated electron systems, including Density Matrix Embedding Theory (DMET), constrained path Auxiliary Field Quantum Monte Carlo (CP-AFQMC), Density Matrix Renormalization Group (DMRG), Variational Monte Carlo (VMC), Multi Reference Protected Hartree Fock (MRPHF), Unrestricted Coupled Cluster Theory including Singles and Doubles (UCCSD) and Dynamical Cluster Approximation (DCA) \cite{PhysRevX.5.041041}.  
The possible systematic error of the SRIKR+ slave-boson method is in the choice of the MF approximation. Here, the renormalization of the hopping amplitudes introduced by KR that allows to map the saddle point solution to the result of the Gutzwiller approximation (which is known to become an exact variational estimate in the limit of infinite dimensions) provides an excellent starting point. This is corroborated by our finding that fluctuation contributions to the free energy calculated within SRIKR+ are small, of the order of percent of the mean field value (from a more formal perspective, within a generalized model the one-loop order expansion used becomes exact in the limit of infinite number of flavors).
Possible systematic errors of the benchmark data may be traced to the finite system cluster sizes for DMET, the constrained path approximation for CP-AFQMC that resolves the sign problem for finite doping, the finite width in the periodic direction for DMRG and the trial wave functions for MRPHF and UCCSD \cite{PhysRevX.5.041041,zheng2017stripe}.
The data shown for comparison in the figures, calculated with the help of DMET, AFQMC, DMRG, MRPHF and DCA has been extrapolated to the thermodynamic limit (TDL), using an ad hoc extrapolation scheme, while in the case of UCCSD data for the largest available cluster sizes has been chosen. In a recent work \cite{Varbenchmark2024} a measure of the systematic error named V-score involved in the application of a given method has been devised. For the Hubbard model at finite doping and methods mentioned above, the estimated relative error is of the order of several times $10^{-3}$. This is of the order of the difference by which our results deviate from state-of-the-art methods mentioned above.

In \autoref{sec:t20} and \autoref{sec:t202} we discuss in detail the results for next-to-nearest neighbor hopping amplitudes $t'=0$ and $t'=\pm0.2$, respectively. Overall, the SRIKR+ method consistently yields the lowest free energy among these methods, whereby the fluctuation corrections contribute a significant improvement over the MF value. The DMET and CP-AFQMC energies are situated between the SRIKR+ and the KRSB MF result and MRPHF and UCCSD are close to the MF energy or higher (see Figures \ref{fig:FEt20} to \ref{fig:n1d}). In the following, we investigate how finite cluster sizes affect the results. 
Notably, the SRIKR+ method does not suffer from finite size effects, i.e., there is no bias towards commensurate order and the calculations are done in the TDL without extrapolation. Our method allows to identify important contributions to the free energy from fluctuations of given momenta. We shall use this information in the following to identify fluctuations present in a large system that are absent in small systems, such as incommensurate fluctuations.
This is especially impactful for the doped Hubbard model, where incommensurate fluctuations are known to be present \cite{PhysRevLett.64.1445,Hubbard_Wuerzburg,PhysRevB.110.085104} as the spin susceptibility in \autoref{fig:susceptability} indicates.

In \autoref{sec:incomm}, we quantify the impact of incommensurate fluctuations, concluding that lattices of at least $8\times 8$ are required to achieve a good approximation of the TDL in our method. Unfortunately, DMET data for such large lattices is not yet available.

\subsection{Short-range hopping model: $t'=0$}\label{sec:t20}

\autoref{fig:FEt20} shows the SRIKR+ free energy in comparison to benchmark data from other methods as a function of the density $n$ for $U=2$ and $U=4$ respectively. The free energy corrected by fluctuations is displayed by the orange line, while the gray line is the MF value. The upper panels show the fluctuation correction relative to the MF energy, which goes to zero for $n\rightarrow 0$ and monotonously increases with the density. The spin fluctuation contribution is found to be approximately twice as large as the charge fluctuation contribution. The data is shown up to the respective critical densities $n_c$ that mark the onset of incommensurate magnetic order, respectively. Fluctuations within the magnetic domain are beyond the scope of the present work. The impact of fluctuation corrections increases strongly with interaction $U$, reaching a maximum of about $7\%$ of the MF value at $U=4$ compared to approximately $2\%$ at $U=2$. For $n=0.8$, the SRIKR+ free energy is about $0.8\%$ lower than its DMET counterpart at $U=2$, and approximately $3.6\%$ lower at $U=4$.

For a given chemical potential $\mu_0$, the fluctuation corrections yield a small density correction $n^{(2)} > 0$, which increases the total density over its MF value $n^{(0)}$, resulting in a slight reduction of the magnetic domain in the $n$-$U$ plane. The densities that are already magnetic according to the MF are denoted by the gray dashed line. The density corrections are further discussed in \autoref{sec:densitycorrections}. Notice that the shown benchmark points from DMET are paramagnetic in agreement with SRIKR+, i.e., show no sign of magnetic or superconducting order.

\subsection{Long-range hopping model: $t'=\pm 0.2$}\label{sec:t202}

\begin{figure*}
    \centering
    \includegraphics[width=0.99\linewidth]{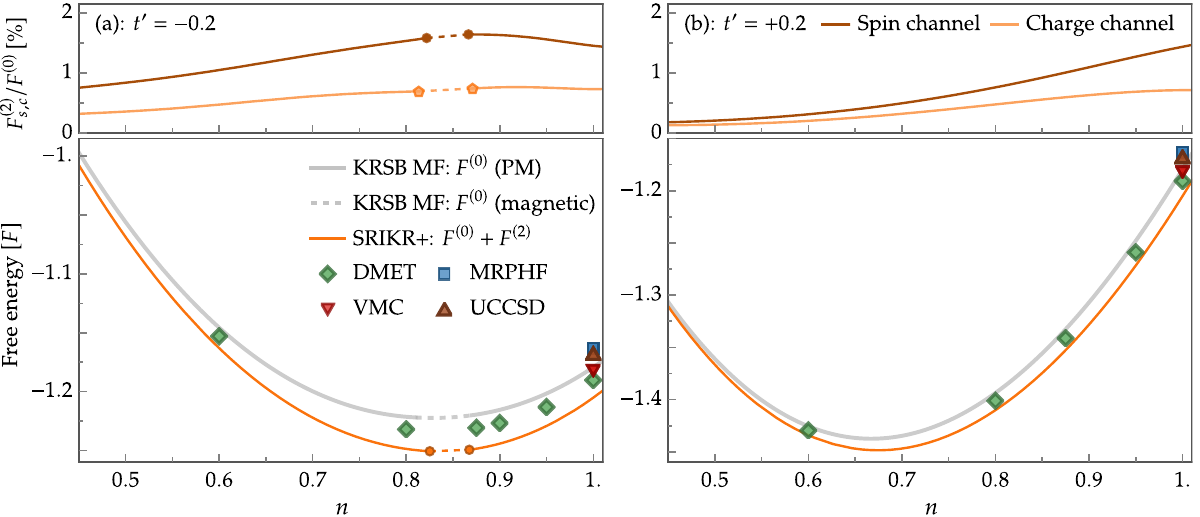}
    \caption{SRIKR+ free energy of the Hubbard model as a function of the density $n$ for $U=2$ and (a): $t'=-0.2$, (b): $t'=+0.2$ along with benchmark data from different methods \cite{PhysRevX.5.041041}. The top panel shows the fluctuation corrections relative to the MF result, which consistently lower the ground state energy.}
        \label{fig:FEU=2}
\end{figure*}

A next-to-nearest neighbor hopping $t'=-0.2$ is of particular interest because the respectively parametrized 2D Hubbard model has been established as an effective minimal model of the high $T_c$-cuprates. In that context, the KRSB approach has previously been applied with great success, achieving quantitative agreement with regard to incommensurate magnetic order for La-based cuprates \cite{PhysRevB.110.085104} and intertwined spin and charge order for the electron-doped cuprate NCCO \cite{Riegler_2023}. Here, we only consider hole-doping but since the SRIKR+ method respects the particle-hole symmetry of the model
\begin{equation}
F(\delta,t',U)=F(-\delta,-t',U)-U\delta   
\end{equation}
with $\delta > 0 $ being hole-doping and $\langle e^2 \rangle -\langle d^2\rangle =\delta$, the results for $t'=0.2$ and hole-doping can be directly mapped to $t'=-0.2$ and electron-doping.

\subsubsection{Free energy}

\autoref{fig:FEU=2} shows the free energy at $U=2$ as a function of doping $\delta = 1-n$ for (a): $t'=-0.2$ and (b): $t'=+0.2$, respectively. Due to particle-hole symmetry, we find the same result at half-filling, but there are substantial differences with finite doping. These are mainly due to the van-Hove singularity, which is situated around $n\approx 0.84$ for $t'=-0.2$ and absent for $t'=+0.2$. Around the singularity, there is a narrow region of incommensurate AFM order of $(\pi,Q)$-type, denoted by the dashed lines, where we do not evaluate fluctuation corrections within the scope of the present work. Apart from that, we detect a stable PM ground-state in agreement with the DMET benchmark. As a consequence of the higher DOS and the closeness to a magnetic phase transition for $t'=-0.2$, we consistently find larger fluctuation corrections in this case. Their magnitude is maximized at the densities that employ the lowest critical interaction toward magnetic order, i.e., $n\approx0.85$ for $t'=-0.2$ and $n=1$ for $t'=+0.2$. 

\subsubsection{Density corrections}\label{sec:densitycorrections}

\begin{figure*}
    \centering
    \includegraphics[width=0.99\linewidth]{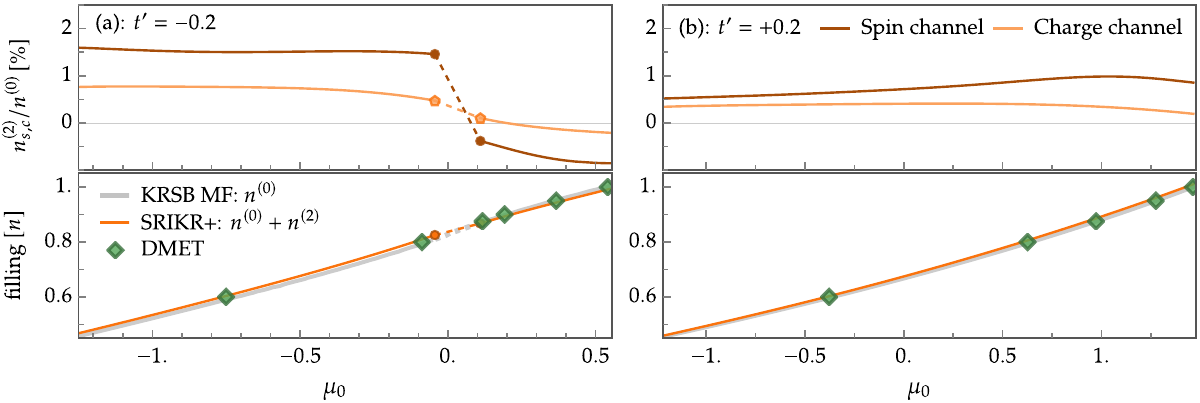}
    \caption{Total density $n_{tot}=n^{(0)}+n^{(2)}$ as a function of the chemical potential $\mu_0$ in the Hubbard model for $U=2$ and (a): $t'=-0.2$, (b): $t'=+0.2$ along with benchmark data from DMET \cite{PhysRevX.5.041041,PhysRevB.93.035126}. The top panel shows the fluctuation corrections relative to the MF result, which consistently lead to a slight decrease of the magnetic domain in the $n-U$ plane.}
        \label{fig:n}
\end{figure*}

Based on the fluctuation corrections of the grand potential for a fixed $\mu_0$, we are considering fluctuation corrections to the density labeled by $n^{(2)}$ according to \eqref{eq:n2}. To display their magnitude, we show the MF and fluctuation-contribution to the density as a function of the chemical potential for $t'=\pm0.2$ in \autoref{fig:n}.
The density corrections dictate small corrections to the magnetic phase boundaries in the $n-U$ plane, whereas they are unchanged in the $\mu_0-U$ plane. This is illustrated in \autoref{fig:FEt20} and \autoref{fig:FEU=2}, where the gray-dashed lines indicate the magnetic MF domains, and the orange-dashed lines indicate the fluctuation-corrected magnetic domains. For $t'=0$ and $t'=+0.2$, the corrections are found to be greater than zero, i.e., push the magnetic phase boundaries closer to half-filling and effectively reduce the size of magnetic domain. For $t'=-0.2$, there is a jump around the magnetically ordered domain, where $n^{(2)}>0$ for small dopings and $n^{(2)}<0$ close to half-filling, which also leads to a slight reduction of the magnetic domain. 

\subsubsection{Double occupancy}\label{sec:resuts:n1}

\begin{figure}
   \includegraphics[width=\linewidth]{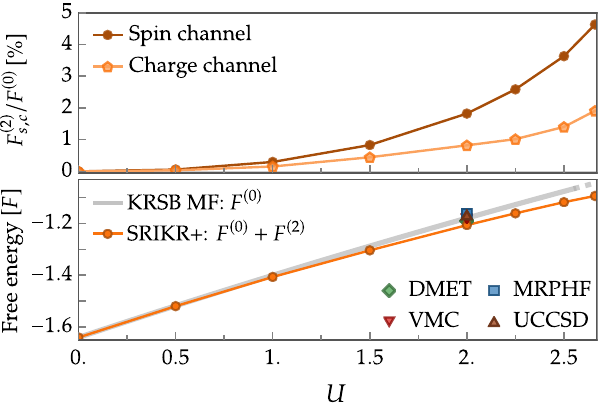}
   \caption{SRIKR+ free energy as a function of the interaction $U$ at $ n=1,t'=-0.2$ along with benchmark data \cite{PhysRevX.5.041041}. The fluctuation corrections consistently vanish at $U=0$.}
      \label{fig:n1U}
\end{figure}

\begin{figure}
        \includegraphics[width=\linewidth]{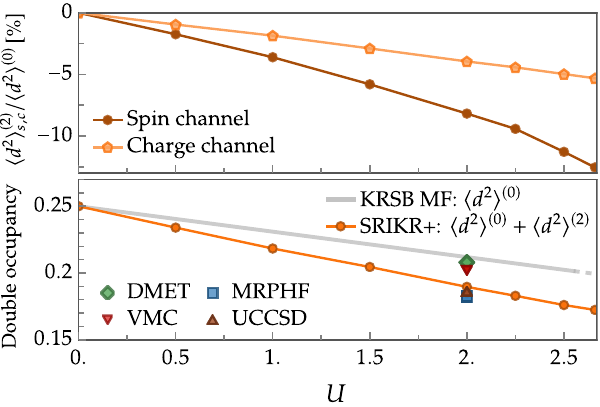}
     \caption{SRIKR+ double occupancy as a function of the interaction $U$ at $ n=1,t'=-0.2$ along with benchmark data \cite{PhysRevX.5.041041,PhysRevB.93.035126}. The fluctuations implicate a reduction of the average number of doubly occupied states.}
   \label{fig:n1d}
\end{figure}

Fluctuation corrections to the average number of doubly occupied states per lattice site are determined by
\begin{equation}
\langle d^2 \rangle = \frac{\partial F }{\partial U} = \langle d^2 \rangle^{(0)} + \frac{\partial F^{(2)}}{\partial U},
\end{equation}
where $\langle d^2 \rangle^{(0)}=\bar{d}^2$ is the MF value and the second term denoted by $\langle d^2 \rangle^{(2)}$ is the fluctuation correction. We show the free energy at half-filling as a function of the interaction in \autoref{fig:n1U} up to the magnetic transition happening at $U\gtrsim 2.6$, where the corrections to the double occupancy in \autoref{fig:n1d} are given by its slope. Notice that we first calculate the grand potential at a fixed chemical potential and from there the corresponding average density including fluctuation corrections in our approach. Calculating the free energy at a given density like $n=1$, requires a fine-tuning of the chemical potential. For the data presented, the density is in the range $n=1.0\pm 0.0025$.
In \autoref{sec:app:U0} we show that the fluctuation corrections vanish in the non-interacting limit ($U=0$) such that the exact result of free fermions is recovered. This is also confirmed by our numerical results, compare \autoref{fig:n1U}. Moreover, we find $F^{(2)}\propto U^2$ for $U\rightarrow0$, such that the correction to the double occupancy also vanishes in this limit. As expected, increasing interactions leads to a reduced number of doubly occupied states, and this trend is further amplified by the fluctuation corrections. Although the SRIKR+ free energy is found to be the lowest among the available benchmark data, the double occupancy lies between MRPHF, UCCSD, and VMC, DMET, where the latter two involve only a small reduction compared to our MF value.

\begin{figure}
    \centering
         \includegraphics[width=\linewidth]{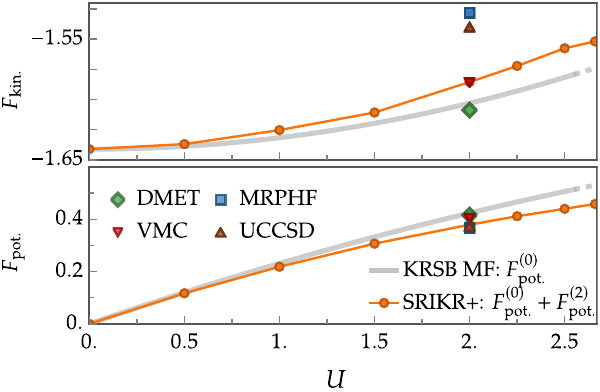}
        \caption{Kinetic and potential energy as a function of the interaction $U$ at $ n=1,t'=-0.2$ along with benchmark data \cite{PhysRevX.5.041041,PhysRevB.93.035126}. The fluctuation contribution decreases the potential energy and increases the kinetic energy compared to the MF value.}
 \label{fig:FEkinpot}
\end{figure}

With the result for $\langle d^2 \rangle^{(2)}$, we can distinguish between corrections to the potential energy $F^{(2)}_{pot.}=U\langle d^2 \rangle^{(2)}$ and the kinetic energy $F^{(2)}_{kin.}=F^{(2)}-F^{(2)}_{pot.}$ in the limit of low temperature. 
We observe the fluctuation corrections to substantially lower the potential energy at the cost of a slightly higher kinetic energy with a net reduction of the total ground state energy. In other words, the effective bandwidth is slightly decreased by fluctuations, which is due to the reduced mobility of the electrons through the avoidance of double occupancies.
At $U=2$, the free energy deviation from SRIKR+ to DMET is $1.03\%$. The deviation of $d^2$ is $8.4\%$, meaning that SRIKR + has a lower potential energy and a higher kinetic energy compared to DMET. In fact, we find the kinetic energy of DMET to be approximately equal to the MF kinetic energy of KRSB. This is demonstrated in \autoref{fig:FEkinpot}.

With the corrections to the double occupancy and density corrections discussed in the previous section, we can also infer corrections to the average occupation of the singly occupied sites $\langle p_0^2\rangle^{(2)} = n^{(2)}-2\langle d^2\rangle^{(2)}$ and of empty sites $\langle e^2\rangle^{(2)} = \langle d^2\rangle^{(2)}-n^{(2)}$. We, therefore, have access to fluctuation contributions to all bosonic occupation numbers.

\subsection{Systems of finite size: impact of incommensurate fluctuations}\label{sec:incomm}

\begin{figure}
    \centering
    \includegraphics[width=0.99\linewidth]{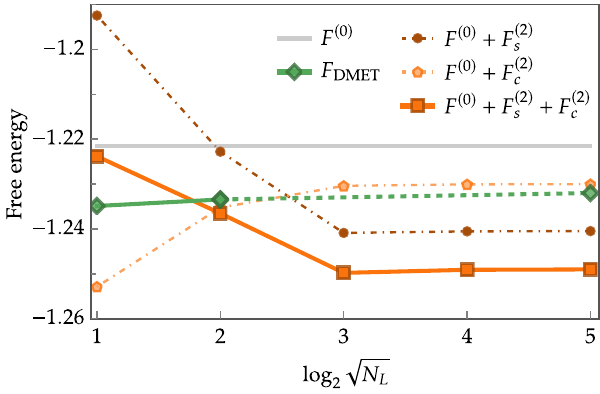}
    \caption{Free energy as a function of $N_L$, the number of sites in the cluster (DMET) or else the fluctuation-lattice size (SRIKR+) for $n=0.8,U=2,t'=-0.2$. The comparison shows SRIKR+ data for coarse-grained fluctuation momenta and DMET benchmark data \cite{PhysRevB.93.035126}. The last DMET point is not an actual cluster result but represents an extrapolation to the TDL.}
    \label{fig:clusters}
\end{figure}

As mentioned above, when comparing our results with benchmark results obtained for small finite-size clusters we consistently find our results to be somewhat lower in energy, compared to the extrapolated thermodynamic limit values presented in \cite{PhysRevX.5.041041}. The effect of cluster size may be probed within our calculations by restricting the discrete momentum values used in summing over fluctuation momenta to small values such as $2\times2$, $4\times4$,.... The restriction to only few ``lattice points" may lead to substantially different results, in particular, in the presence of incommensurate fluctuations, which may not be well captured in the coarse-grained approximation. 
As a general trend, we find that the free energy is lowered with increasing number $N_L$ of ``lattice points". This observation calls in question the simple extrapolation scheme used in \cite{PhysRevX.5.041041}, which continued the trend of the free energy found there to weakly increase with $N_L$ for small cluster sizes up to infinite cluster size $N_L \rightarrow \infty$, which clearly contradicts our findings as shown in  \autoref{fig:clusters}

\begin{figure}
    \centering
    \includegraphics[width=0.99\linewidth]{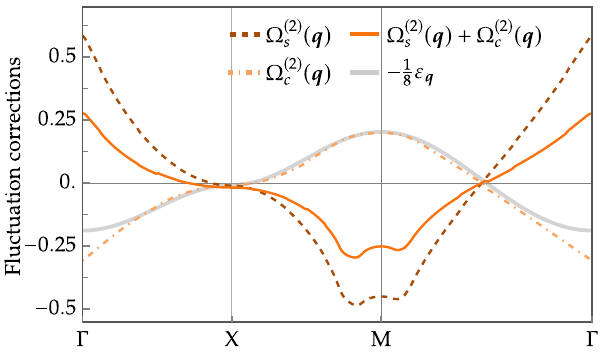}
    \caption{Momentum-resolved fluctuation contributions of the spin and charge channel for $n=0.8,U=2,t'=-0.2$.}
    \label{fig:OmegaQ}
\end{figure}

To quantify the impact of incommensurate fluctuations, we calculated the momentum-resolved contributions of the fluctuation corrections $\Omega^{(2)}_{s/c}(\boldsymbol q)$, where
\begin{equation}
    \Omega^{(2)}_{s/c} = \frac{1}{N_L} \sum_{\bs q \in \text{BZ}} \Omega^{(2)}_{s/c}(\boldsymbol q).
\end{equation}
The result for $n=0.8,U=2,t'=-0.2$ is shown in \autoref{fig:OmegaQ} on the HSL of the BZ, where the spin and charge channel are displayed by dashed lines and the combined result by the solid line. The approximate total fluctuation contribution is given by the area under the curve. Its momentum dependence is mainly determined by the characteristic energy
\begin{equation}
    \varepsilon_\bq = -\frac{2}{N_L} \sum_\bk n_F(\epsilon_\bk)\xi_{\bk+\bq}
\end{equation}
and the susceptibilities $\chi_m(\bq,i\omega_n)$ which are coupled in the fluctuation determinant and through the Matsubara summation, see \app{app:sec:fluctuationmatrix}. The charge sector in \autoref{fig:OmegaQ} mostly follows the momentum-dependence of $-\varepsilon_\bq$, while characteristic features of the charge susceptibility $\chi_c$ are absent. This is different in the spin-sector, where the momentum-dependence somewhat follows $\varepsilon_\bq$ but additional characteristic incommensurate features in analogy to the spin-susceptibility $\chi_s$ are present near the $\mathrm{M}$-point. These incommensurate contributions are not captured on small lattices.

This is demonstrated in \autoref{fig:clusters}, which shows the free energy as a function of the fluctuation-lattice size, i.e., number of bosonic momenta. Since the PM MF free energy does not contain incommensurate contributions, the MF lattice is kept in the TDL. For small lattices, the impact of the $\Gamma$ and $\mathrm{M}$-point is enhanced, yielding an overestimation of the charge-sector and an underestimation of the spin-sector with overall higher energies. The SRIKR+ free energy result at $N_L=16$ is quite close to the DMET benchmark data, whereas the extrapolated DMET data point, here plotted at $N_L=1024$, is definitely too high. Convergence of the SRIKR+ data is only achieved for lattices of $8\times 8$ sites and larger.
A sign of incommensurate fluctuations is also visible in the DMET data, where the $2\times 8$ cluster yields a very small magnetization that vanishes in the $4\times 4$ cluster for $n=0.8,t'=-0.2,U=2$ \cite{PhysRevB.93.035126}. At the same parameters except for higher interaction $U=6$ it has previously been shown that the incommensurate magnetic order found with SRIKR matches the DMET result since the spin ordering vector matches a periodicity of $8$ lattice sites at this doping \cite{Hubbard_Wuerzburg}. We emphasize that the impact of incommensurate fluctuations is maximized close to magnetic phase transitions and finite doping as in the case presented. 

%%---------------------------------------------------------------------
%%---------------------------------------------------------------------
%% RESULTS AT T>0
%%---------------------------------------------------------------------
%%---------------------------------------------------------------------

\section{Results at finite temperature}
 We aim to determine the contribution of fluctuations to the quasi-particle effective mass. Within the usual quantum many-body formalism these contributions are accessible via the leading low-frequency term of the self-energy, which contributes to the leading $T^2$ law of the free energy. The direct contribution of fluctuations to the free energy leads to a $T^3 \ln{T}$ term \cite{PhysRev.169.417} (here the inclusion of transversal spin fluctuations as done by the SRIKR method is required \cite{wolfle1990spin}) , while the $T^2$ terms contributed by the fluctuations directly may be shown to cancel \cite{brenig1967specific}, as expected in Fermi liquid theory. In the present slave boson calculation the $T^2$ contribution of fluctuations to the free energy may be calculated directly (it is not known how to define something like the electron self-energy within this method). In other words, the $T^2$-contribution to the free energy calculated below corresponds to the self-energy calculation within a conventional many-body calculation.

 The leading-order temperature-dependence in the SRIKR+ approach is found to be
 \begin{equation}
 F(T)= \left(F^{(0)} + F^{(2)}\right)\Big{|}_{T=0} - \frac{1}{2} \gamma T^2 +... 
 \end{equation}
where $\gamma=\gamma^{(0)}+\gamma^{(2)}$. We can thus calculate fluctuation corrections to the temperature coefficient $\gamma$ of the specific heat
 \begin{equation}
 C(T)= - T \frac{\partial^2 F}{\partial T^2} = \gamma T 
 \end{equation}
The effective mass ratio is defined as
\begin{equation}
 m^*/m= \gamma/\gamma_{0} 
 \end{equation}
where $\gamma_{0}=\lim_{U \rightarrow 0} \gamma$ is the specific heat coefficient in the non-interacting limit. 

\subsection{Interaction-dependence at half-filling}

\begin{figure}
    \centering
    \includegraphics[width=0.99\linewidth]{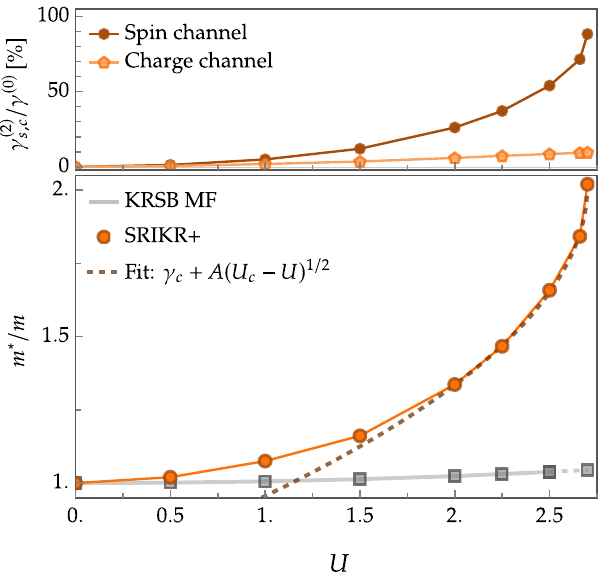}
    \caption{Effective mass ratio $m^*/m$ including fluctuation corrections as a function of $U$ at half-filling and $t'=-0.2$. The steep increase of the effective mass around the magnetic transition at $U_c=2.7$ is fitted with \eqref{eq:mstar} and $\gamma_c=2.01, A=-0.81$. The top panel shows the fluctuation contributions relative to the MF value and illustrates that the enhancement is mainly due to spin fluctuations.}
    \label{fig:Gamma}
\end{figure}

\autoref{fig:Gamma} shows the effective mass ratio $m^*/m$ as a function of the interaction at half-filling. While the MF result only leads to an increase of effective mass of approx $3.8\%$ at the magnetic transition, the fluctuation contribution $\gamma^{(2)}$ yields an increase of more than $100\%$. The increase of the effective mass around $U_c\approx2.7$ follows the behavior
\begin{equation} \label{eq:mstar}
\frac{m^*}{m} = \gamma_{c} + A \left(U_c-U \right)^{1/2},  
\end{equation}
where $U_c$ marks the critical interaction, where $\chi_s(\omega=0,\bq=\mathrm{M})$ diverges. The onset of AFM is a first-order phase transition at $U=2.7$, such that the PM state is meta-stable for $2.7 \leq U \leq 2.8$. These numbers include fluctuation corrections to the density. On MF level, the magnetic transition is at $U=2.55$ as indicated by the gray dashed line in \autoref{fig:Gamma}.

It is striking that the spin-fluctuation contributions to $\gamma$ are much larger than the charge-fluctuation contributions, despite being of the same order of magnitude at $T=0$. This is because the coefficient $\gamma$ probes the low-energy thermal excitations, whereas the ground state energy is renormalized by quantum fluctuations of all energy scales. The charge-susceptibility at moderate interaction $U$ is not sufficiently enhanced in contrast to the spin-susceptibility. Consequently, the charge-fluctuations are mainly determined by the function $\varepsilon_\bq$, which features a much lower temperature-dependence than the Lindhard function that enters the susceptibilities, compare \autoref{sec:incomm}.

\subsection{Electron-doping}

\begin{figure}
    \centering
    \includegraphics[width=0.99\linewidth]{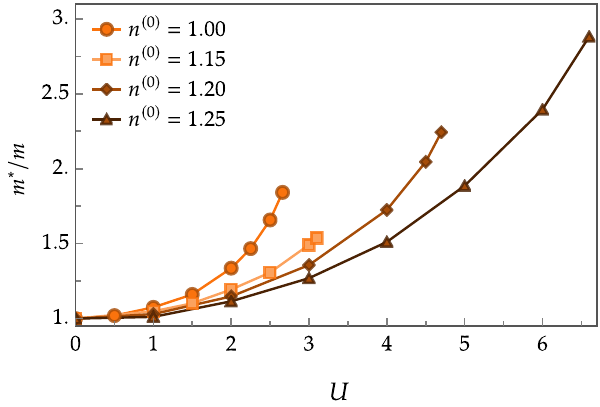}
    \caption{Effective mass as a function of $U$ for $t'=-0.2$ and different dopings. The data points are shown up to the critical $U_c$ that marks the onset of magnetic order at each density respectively. The specific heat coefficient at $U=0$, i.e., $\gamma_0$ for the shown densities $n^{(0)}=1.00,1.15,1.20,1.25$ is $1.35,1.03,0.96,0.90$, respectively.}
    \label{fig:GammaN}
\end{figure}

\autoref{fig:GammaN} shows the effective mass as a function of $U$ for different values of electron-doping up to the respective critical $U_c$ that implicates magnetic order. The steepest increase in effective mass is found at zero doping, but since $U_c$ increases with doping, the highest possible values of $m^*/m$ are found for higher doping.

%%---------------------------------------------------------------------
%%---------------------------------------------------------------------
%% CONCLUSION
%%---------------------------------------------------------------------
%%---------------------------------------------------------------------

\section{Conclusion}
In the present work we have shown that the spin-rotation invariant Kotliar-Ruckenstein slave-boson method including High Frequency Contributions (SRIKR+) allows calculating thermodynamic and dynamic properties of strongly correlated electron systems with very good accuracy, competitive with the best available methods. It has been known for some time that the mean-field results of SRIKR, which are identical to the Gutzwiller approximation, are of good quality to begin with. In the present work, we evaluated the contribution of fluctuations about the mean field in the paramagnetic state. One may distinguish charge fluctuations and spin fluctuations. The spin-rotation-invariant formulation allows to capture both, longitudinal and transverse spin fluctuations. The latter are missing in the initial KR formulation.

The controlled calculation of the contribution of fluctuations to the free energy within the KR slave boson method has been an unsolved problem for three decades. The problem is rooted in the usual practice of taking the limit of continuous imaginary time in the functional integral representation of the grand partition function from the outset, rather than at the end of the calculation in the discrete-time mode. In the present work, we succeeded in resolving these problems with the help of insights gained by Arrigoni and Strinati \cite{PhysRevB.52.2428} in the early nineties. At the core of our work is the recognition of the important role of the ordering of operators in the Gutzwiller renormalized kinetic energy operator proposed by Kotliar and Ruckenstein. This ordering is essential when calculating ``high-frequency contributions" (HFC) to the continuous-time functional integral evaluation of the grand partition function. As noted by several authors, the results of the naive CT evaluation of the free energy are of the order of the mean-field value, which by itself would render the fluctuation calculation useless \cite{PhysRevB.44.2403,arrigoni1994functional}. In addition, the CT result strongly violates the non-interacting limit. Applying the HFC leads to a reduction of the magnitude of the fluctuation-contribution to the free energy by two orders of magnitude. The results are now in very good agreement with benchmark data, where available.

The response functions are not affected by the HFC, since their contribution to the generating functional is independent of the generating fields. This means that the many works in the past reporting the calculation of charge or spin response functions remain valid \cite{JWRasul_1988,PhysRevB.41.142,Li1991,woelfle_spin_1997}.

Our starting point into this project has actually been to calculate the contribution of fluctuations to the effective quasiparticle mass. In standard perturbation theory, the effective mass ratio $m^*/m$ is determined by the self-energy. The contribution of fluctuations to the low temperature free energy $F(T)$ is of the form $T^3\ln{T}$, whereas the effective mass shows up in the $F(T) \propto (m^*/m) T^2$ term. The situation is different in the SRIKR+ formulation: one starts with a mean field result for $m^*/m$ , which already incorporates important interaction effects. The effect of fluctuations is not encoded in the self-energy, because there is no such concept in the KR slave boson formalism, but instead the fluctuations add a $T^2$-term to the free energy, thus modifying the effective mass. We have determined this modification and found it to be substantial in situations where fluctuations are large. One such situation is in the phase diagram near a transition into a magnetically ordered state where the corresponding Fourier component of the spin fluctuations diverges. This in turn is showing up in a sharp increase in the effective mass.

We demonstrated in this paper that the SRIKR+ method is a powerful tool to calculate thermodynamic and dynamic properties of strongly correlated electron systems. Further applications to magnetically ordered states of the Hubbard model, to the periodic Anderson model and more are in preparation.

%%%%%%%%%%%%%%%%%%%%%%%%%%%%%%%%%
%         Acknowledgments       %
%%%%%%%%%%%%%%%%%%%%%%%%%%%%%%%%%
 
\begin{acknowledgments}
	\textit{Acknowledgments}--- 
	The authors thank Jörg Schmalian, Raymond Fr\'esard, Thilo Kopp and Werner Hanke for helpful discussions. The work at KIT is funded by Deutsche Forschungsgemeinschaft (DFG, German Research Foundation) Project No. SCH 1031/13-1. The work in Würzburg is funded by Deutsche Forschungsgemeinschaft (DFG, German Research Foundation), Project No. 258499086 - SFB 1170 and the Würzburg-Dresden Cluster of Excellence on Complexity and Topology in Quantum Matter \textit{ct.qmat} Project No. 390858490 - EXC 2147. Peter W\"olfle acknowledges support through a Distinguished Senior Fellowship of Karlsruhe Institute of Technology. 
\end{acknowledgments}
	
%%%%%%%%%%%%%%%%%%%%%%%%%%%%%%%%%
%          Appendix         %
%%%%%%%%%%%%%%%%%%%%%%%%%%%%%%%%%

\appendix

\section{Mean field solution}\label{app:sec:MF}
The mean field solution of the slave boson amplitudes and of the constraint fields will be denoted by an overbar. It is useful to employ the variable $x=\overline{e}+\overline{d}$, which may be determined analytically as \cite{PhysRevB.35.6703,woelfle_spin_1992}
\begin{align}
x^{2}  &  =\frac{1}{3}(1-u)+\frac{2^{1/3}}{3}\frac{(u-1)^{2}}{\{a_{x}%
+[a_{x}^{2}-4(u-1)^{6}]^{1/2}\}^{1/3}}\\
&\phantom{=}  +\frac{2^{-1/3}}{3} \left\{ a_{x}+[a_{x}^{2}-4(u-1)^{6}]^{1/2}\right\}^{1/3},\nonumber\\
a_{x}  &  =2-6u+27u\delta^{2}+6u^{2}-2u^{3}.
\end{align}
Here, $\delta=1-n$ denotes the hole doping and $u=U/U_{c}$ is the normalized interaction energy, with $U_{c}$ being the interaction at which a Mott metal-insulator transition takes place at $\delta=0$, defined by
\begin{equation}
U_{c}%
=8(1-\delta^{2})^{-1}\varepsilon_0,
\end{equation}
where $\varepsilon_0$ has been defined in \eqref{eq:epsilon0}. We note that $\varepsilon_0$ depends on the chemical potential $\mu_{0}$, wich in turn is related to $\delta$ by 
\begin{equation}\label{eq:delta}
1-\delta = -\frac{2}{N_{L}}\sum_{\bk} n_F(\epsilon_\bk). 
\end{equation}
Thus the above solution requires a self-consistent determination of $\mu_{0}$ for given $\delta$.
In terms of $x$, the occupation numbers and constraint parameters are given by
\begin{align}\label{eq:MFsolution}
\begin{split}
\overline{d}^{2}  &  =\frac{(x^{2}-\delta)^{2}}{4x^{2}},\\
\overline{p}_{0}^{2}  &  =1-\frac{x^{4}+\delta^{2}}{2x^{2}},\\
\overline{\alpha}  &  =\frac{U_{c}}{2}x^{2}\overline{p}_{0}^{2}\left(\frac{1}{1-\delta}
+\frac{1}{x^{2}+\delta}\right),\\
\overline{\beta}_{0}  &  =\overline{\alpha}- \frac{U_{c}}{4}x^{2}\left(1+\frac{2 \overline{p}_{0}^{2}}{1-\delta^2}\right).
\end{split}
\end{align}
The saddle point equations can also conveniently be solved numerically with the the introduction of an effective chemical potential $\mu_{\text{eff}}=\mu_0-\beta_0$ to reduce the number of MF parameters to a minimal set of independent variables. This way, the density $n$ serves as an input parameter and can be set directly, rather than iterating for different $\mu_0$ to obtain the desired density \cite{Hubbard_Wuerzburg}.

\section{Fluctuation matrix in the continuous time (CT) limit}\label{app:sec:fluctuationmatrix}

Within the scope of Gaussian fluctuations, we expand the action up to the second order in bosonic fields around the MF saddle point with the MF action $\mathcal{S}^{(0)}$:
\begin{equation}
 \mathcal{S} = \mathcal{S}^{(0)} + \frac{T}{N_{L}}\sum_{\i\omega_n}\sum_{\bq} \delta\psi_{\mu,-q} \mathcal{M}_{\mu\nu}(q) \psi_{\nu,q} +...    
\end{equation}
The fluctuation matrix is given by
\begin{equation}
\mathcal{M}_{\mu\nu}(q) = \frac12 \frac{\delta ^2 \mathcal{S}}{\delta \psi_{\mu,-q} \delta \psi_{\nu,q}},
\end{equation}
where $q=(i\omega_n,\bf{q})$ and $\omega_n = 2\pi n T$ is a bosonic Matsubara frequency. The evaluation of the fluctuation matrix $\mathcal{M}_{\mu\nu}(q)$ around PM saddle points in the CT limit has been established in the literature and applied to calculate the $T^3\log T$ spin fluctuation contribution to the specific heat \cite{wolfle1990spin} as well as the dynamic spin and charge susceptibility \cite{Li1991,woelfle_spin_1997,PhysRevB.95.165127,Hubbard_Wuerzburg,Riegler_2023}. The approach was also generalized to encompass magnetic saddle points \cite{Seufert_2021}. In the following, we provide the result for $\mathcal{M}_{\mu\nu}(q)$ in agreement with the established literature.

In the fluctuation basis, $\delta\psi_{\mu,q} = \psi_{\mu,q} - \mf\psi_\mu \delta_{q,0}$ describes the deviation from the MF value $\mf \psi_\mu$ and the index $\mu$ labels the fluctuation fields. We apply the radial gauge, where the phases of all fields except for the $d$-field are gauged away such that they are radial fields. Moreover, we decompose the remaining complex $d=d_1+\i d_2$ field into its real part $d_1$ and imaginary part $d_2$. It has been established in Ref.~\cite{Hubbard_Wuerzburg} that an arbitrary slave-boson field can be eliminated from the fluctuation basis by direct application of the occupation constraint
\begin{equation}\label{eq:app:alphaconstraint}
    e^\dagger e^\nodag + p_0^\dagger p^\nodag_{0} + \bs p^\dagger \cdot \bs p^\nodag + d^\dagger d^\nodag = 1
\end{equation}
that is enforced by the Lagrange multiplier $\alpha$, whereby the result for the susceptibilities remains unchanged. With this substitution, the determinant of the fluctuation matrix only changes by a constant factor such that the fluctuation contribution to the grand potential also remains invariant. In the following, we choose to substitute the $e$-field without loss of generality such that $e$ and $\alpha$ drop out of the fluctuation basis, which significantly simplifies the calculation. For fluctuations around PM saddle points, the fluctuation matrix is block-diagonal such that the charge sector with the basis $\delta \bs \psi_c = (d_1,d_2,p_0,\beta_0)^\top$ and the three identical spin sectors spanned by the fields $\delta \bs \psi_{s,\mu} = (p_{\mu},\beta_{\mu})^\top$, with $\mu=1,2,3$ are decoupled. Consequently the determinant of the fluctuation matrix is given by
\begin{equation}
    \det\mathcal{M} = \det\mathcal{M}_c [\det\mathcal{M}_s]^3.
\end{equation}

The results are in full agreement with previous works \cite{Hubbard_Wuerzburg,Seufert_2021,Riegler_2023} but we adopt a more compact notation suggested in Refs.~\cite{woelfle_spin_1997,PhysRevB.95.165127}, which is especially handy for the analytic evaluation of the non-interacting limit that we present in \app{sec:app:U0}. To do so, we define the following quantities
\begingroup
\allowdisplaybreaks
\begin{align}
\xi_\bk &= -2t\left(\cos k_x + \cos k_y\right) - 4t' \cos k_x \cos k_y, \\ 
\epsilon_\bk &=  |\mf z|^2 \xi_{\bk} + \beta_0 -\mu_0, \\
 \varepsilon_0 &= -\frac{2}{N_{L}}\sum_{\bk} n_F(\epsilon_\bk)\xi_\bk, \\
  \varepsilon_\bq &= -\frac{2}{N_{L}}\sum_{\bk} n_F(\epsilon_\bk)\xi_{\bk+\bq}, \\
 \chi_m &= - \frac{2}{N_{L}}\sum_{\bk} \frac{n_F(\epsilon_\bk)-n_F(\epsilon_{\bk+\bq})}{\i\omega_n + \epsilon_\bk-\epsilon_{\bk+\bq}}\left(\xi_\bk + \xi_{\bk+\bq} \right)^m.
\end{align}
\endgroup
The $z$-factor is expanded in Pauli matrices, including the identity matrix $\tau_0$
\begin{equation}
\underline{z} = z_{+} \underline{\tau}_0 + z_{-} \frac{\bs p}{|\bs p|}  \underline{\bs \tau},
\end{equation}
where we define the scalar quantities
\begingroup
\allowdisplaybreaks
\begin{align}
	z_{+} &= \frac{z_\uparrow + z_\downarrow}{2}, \label{eq:app:z_plus}\\
    z_{-} &= \frac{z_\uparrow - z_\downarrow}{2}\label{eq:app:z_minus},\\
	z_{\uparrow,\downarrow} &= \frac{p_0(e+d_1+\i d_2) \pm |\bs p|(e-d_1-\i d_2)}{\sqrt{2(1-d_1^2-d_2^2(p_0\pm |\bs p|)^2/2)(1-e^2-(p_0\mp |\bs p|)^2/2)}}, \\
\overline{z} &= \frac{\overline{p}_0(\overline{e}+\overline{d})}{\sqrt{2(1-\overline{d}^2-\overline{p}_0^2/2)(1-\overline{e}^2-\overline{p}_0^2/2)}},
\end{align}
whereby $\overline{z}$ is the mean field limit of $z$. 
\endgroup
Moreover, we define the derivatives of the $z$-factors
\begingroup
\allowdisplaybreaks
\begin{align}
    \tilde{z}_{\pm,\mu} &= \frac{\partial}{\partial \psi_\mu} z_\pm \left[e=\sqrt{1-p_0^2-\bs p ^2 -d_1^2-d_2^2},p_0,\bs p, d_1, d_2\right],\\
    \tilde{z}_{\pm,\mu\nu} &= \frac{\partial^2}{\partial \psi_\mu\partial \psi_\nu} z_\pm \left[e=\sqrt{1-p_0^2-\bs p ^2 -d_1^2-d_2^2},p_0,\bs p, d_1, d_2\right],
\end{align}
\endgroup
where the $e$-field has been substituted with the constraint in \eqref{eq:app:alphaconstraint} prior to taking the respective derivative (denoted by the tilde) and evaluated at the MF saddle point $\mf\psi$. 

\subsection{Spin sector}

The fluctuation matrix for the spin-sector spanned by the fields $\delta \bs \psi_{s,1} = (p_{1},\beta_{1})^\top$ is given by
\begin{equation}\label{eq:Ms}
    \mathcal{M}_s(q) = 
    \begin{pmatrix}
    -\beta_0 + \mathcal{M}^f_{p_1,p_1}(q)  & -\bar{p}_0 -\dfrac{\bar{z} \tilde{z}_{-,p_1}}{2} \chi_1(q)\\
    -\bar{p}_0-\dfrac{\bar{z} \tilde{z}_{-, p_1}}{2} \chi_1(q) & \dfrac12\chi_0(q)
    \end{pmatrix},
\end{equation}
with 
\begin{equation}
    \mathcal{M}^f_{p_1,p_1}(q) = -\varepsilon_0 \bar{z} \tilde{z}_{+,p_1p_1}- \left(\varepsilon_\bq +\frac12 \mf z^2 \,\chi_2(q) \right) \tilde{z}^2_{-, p_1}.
\end{equation}
Notice that the first derivative $\tilde{z}_{+,p_1}$ vanishes at the saddle point but $\tilde{z}_{-, p_1}$ is non-zero. Moreover, even though we have eliminated the Lagrange parameter $\alpha$ from the fluctuation basis, the result still depends on its MF value since $\mf \alpha\propto \varepsilon_0 $, compare \eqref{eq:MFsolution}.

\subsection{Charge sector}
The result for the charge-sector is more involved because it contains the complex $d$-field:
%-------------------------Mc--------------------------------
\begin{widetext}
\begin{equation}\label{eq:Mc}
\mathcal{M}_c (q)
=
\begin{pmatrix}
U-2\beta_0 +\mathcal{M}^f_{d_1,d_1}(q)  &
\omega_n\left(1 - \dfrac{\tilde{z}_{d_1}\tilde{z}_{d_2}}{2}i\chi_1(q)  \right)  &
\mathcal{M}^f_{d_1, p_0}(q) & 
-2d - \dfrac{\overline{z}_0\tilde{z}_{d_1}}{2}\chi_1(q)\\
%-----------------------------------------------------------
-\omega_n\left(1 - \dfrac{\tilde{z}_{d_1}\tilde{z}_{d_2}}{2} i\chi_1(q)  \right)	&
U-2\beta_0 -\varepsilon_0 \left( \overline{z}_0\tilde{z}_{d_2d_2} + |\tilde{z}_{d_2}|^2 \right)+ \dfrac{|\tilde{z}_{d_2}|^2}{2 \overline{z}^2_0} \omega_n^2\, \chi_0(q)& 
 \dfrac{\tilde{z}_{d_2}\tilde{z}_{p_0}}{2}i\omega_n\,\chi_1(q) &
i\omega_n\,\chi_0(q) \dfrac{\tilde{z}_{d_2}}{2 \overline{z}_0}\\
%-----------------------------------------------------------
\mathcal{M}^f_{p_0,d_1}(q) &
 -\dfrac{\tilde{z}_{d_2}\tilde{z}_{p_0}}{2}i\omega_n\,\chi_1(q) &
 -\beta_0+\mathcal{M}^f_{p_0,p_0}(q) &
 -p_0 - \chi_1(q) \dfrac{ \overline{z}_0 \tilde{z}_{p_0}}{2}\\
%-----------------------------------------------------------
-2d - \dfrac{\overline{z}_0\tilde{z}_{d_1}}{2}\chi_1(q) & -i\omega_n\,\chi_0(q) \dfrac{\tilde{z}_{d_2}}{2 \overline{z}_0} & -p_0 - \chi_1(q) \dfrac{ \overline{z}_0 \tilde{z}_{p_0}}{2} & -\dfrac12\chi_0(q)
\end{pmatrix}, 
\end{equation}
\end{widetext}
where we defined 
\begin{equation}
\mathcal{M}^f_{\mu\nu}(q)=  -\varepsilon_0 \mf z_0 \tilde{z}_{\mu\nu} - \left(\varepsilon_\bq +\frac12 \mf z_0^2 \,\chi_2(q) \right) \tilde{z}_\mu \tilde{z}_\nu,
\end{equation}
and where $\mu,\nu=(d_{1},d_{2},p_{0},\beta_{0})$.
For simplicity of notation, we have omitted the $\pm$ subindex of the $z$-factor derivatives, since $z_-=0$ for the charge sector, i.e. all derivatives are those of ~$z_+$ here.
\eqref{eq:Mc} reproduces the result given in~\cite{PhysRevB.95.165127} when replacing $z\rightarrow \tilde{z}$ and removing the $\mf\alpha$ terms due to our substitution, where $e,\alpha$ are not part of the fluctuation basis. This reduces the dimension of the matrix from $6\times 6$ to $4\times 4$ while recovering the same results for the observables, i.e., susceptibilities and fluctuation corrections.
Notice that the matrix elements associated with the $d_2$-field can be expressed in such a way that they do not depend on $\varepsilon_\bq$ or $\chi_2(q)$ such that their momentum dependence is fully contained in $\chi_0(q),\chi_1(q)$. 

\section{Generating functional}\label{app:sec:functionslintegrals}

In this section, we derive the generating functional of the correlation functions. As already mentioned, in transforming the initial field amplitudes from position-time space to Fourier space, e.g., $d_{\i}(\tau)\rightarrow d_{q}$, where $q=(i\omega_{n},\bf{q})$, the number of real-valued fields in position-time space is doubled in Fourier space since the Fourier transformed fields are complex-valued. However, only half of the fields are independent since $d_{-q}=d_{q}^\ast$. This is corrected by taking the square root of the product of Fourier components.

\subsection{Spin sector}
The generating functional for the spin sector spanned by the fluctuation amplitudes $(p_{1},\beta_{1})$ is defined by
\begin{equation}
Y_{fl,cont}^{(s,1)}(J_{p},J_{\beta})=c\left(  \prod_{q}Y_{q}^{(s,1)}
(J_{p},J_{\beta})\right)^{1/2},
\end{equation}
where
\begin{widetext}
\begin{align}
\begin{split}
Y_{q}^{(s,1)}(J_{p},J_{\beta})  & =\int_{0}^{\infty}d(p_{1}^{\ast}p_{1}%
)\int_{-\infty}^{\infty}d\beta_{1}\exp\left\{-[p_{1}^{\ast}p_{1}M_{pp}+(\beta
_{1}^{\ast}p_{1}+h.c)M_{p\beta}+\beta_{1}^{\ast}\beta_{1}M_{\beta\beta
}]+(p_{1}^{\ast}J_{p}+\beta_{1}^{\ast}J_{\beta}+h.c.)\right\}\\
& =\int_{0}^{\infty}d(p_{1}^{\ast}p_{1})\exp\left\{-[p_{1}^{\ast}p_{1}M_{pp}%
+(p_{1}^{\ast}J_{p}+h.c.)\right\}I(p_{1}^{\ast},p_{1}).
\end{split}
\end{align}
\end{widetext}
Here, $M_{pp},M_{p,\beta},M_{\beta,\beta}$ are the matrix elements of the matrix $M_{s}$ given in \eqref{eq:Ms}. The integration of $\beta_{1}$ may be done first, resulting in
\begin{widetext}
\begin{align}
I(p_{1}^{\ast},p_{1})  & =\int_{-\infty}^{\infty}d\beta_{1}\exp\left\{-[(\beta
_{1}^{\ast}p_{1}+p_{1}^{\ast}\beta_{1})M_{p\beta}+\beta_{1}^{\ast}\beta
_{1}M_{\beta\beta}]+(\beta_{1}^{\ast}J_{\beta}+h.c.)\right\}\\
& =\frac{\pi}{|M_{\beta\beta}|}\exp\left\{M_{\beta\beta}^{-1}(M_{p\beta}p_{1}%
^{\ast}-J_{\beta}^{\ast})(M_{p\beta}p_{1}-J_{\beta})\right\}.\nonumber
\end{align}
\end{widetext}
To proceed with the remaining integral of $p_{1}$, we decompose the
complex-valued fields into real and imaginary parts, $p_{1}=p_{1}^{\prime
}+ip_{1}^{\prime\prime}$, etc. and express the integration over the radial
field $p_{1}$ as $\int_{0}^{\infty}2\pi|p_{1}|d|p_{1}|..=\int_{-\infty
}^{\infty}dp_{1}^{\prime}dp_{1}^{\prime\prime}..$, leading to
\begin{widetext}
\begin{align}
\begin{split}
Y_{q}^{(s,1)}(J_{p},J_{\beta})  & =\frac{\pi}{|M_{\beta\beta}|}\int_{-\infty
}^{\infty}dp_{1}^{\prime}dp_{1}^{\prime\prime}\exp\left\{-\left[\left(p_{1}^{\prime2}+p_{1}^{\prime\prime2}\right)M_{D}+2p_{1}^{\prime}J_{D}^{^{\prime}}+2p_{1}%
^{\prime\prime}J_{D}^{^{\prime\prime}}-M_{\beta\beta}^{-1}J_{\beta}^{\ast
}J_{\beta}\right]\right\}\\
& =\frac{\pi^{2}}{|\det M_{s}(q)|}\exp\left\{M_{\beta\beta}^{-1}J_{\beta}^{\ast
}J_{\beta}+M_{D}^{-1}J_{p}^{\ast}J_{p}\right\},
\end{split}
\end{align}
\end{widetext}
where we defined $M_{D}=M_{pp}-M_{\beta\beta}^{-1}M_{\beta p}^{2}$ , with
$J_{D}=J_{p}+M_{\beta\beta}^{-1}M_{\beta p}J_{\beta}$\ and used that
$|M_{\beta\beta}||M_{D}|=|\det M_{s}(q)|$.\ The correlation function $\chi_{pp}(q)$  follows as
\begin{align}
\begin{split}
\chi_{pp}(q) &  =\frac{\partial^{2}}{\partial J_{p,q}^{\ast}\partial J_{p,q}}\ln
Y_{fl,cont}^{(s,1)}\left(\{J_{p},J_{\beta}\}\right)\Big{|}_{J=0}\\
&  = M_{D}^{-1}= \frac{M_{\beta\beta}}{\det M_s}= M^{-1}_{pp}(q).
\end{split}
\end{align}
The spin susceptibility is obtained as 
\begin{equation}
\chi_{s}(q)=2 \overline{p}_{0}^2 \chi_{pp}(q).
\end{equation}
This result is not changed by HFC since the
corresponding corrections do not depend on the generating fields. The
contribution to the grand partition function in the continuous time limit is obtained as
\begin{equation}
Z_{fl,cont}^{(s,1)}=Y_{fl,cont}^{(s,1)}(0,0)=\prod_{q}\sqrt{\frac{\pi}{|\det
M_{s}(q)|}}.
\end{equation}

\subsection{Charge sector}
We choose the fluctuation vector
$\Phi_{cq}=(d_{1q},d_{2q},p_{0q},\beta_{0q})^{\top}$ (after eliminating the field $e_{q}$) and observe $\Phi_{c,-q}=\Phi_{cq}^{\ast}$.  
The fact that the fluctuation matrix $\mathcal{M}_{c}(q)$ given by \eqref{eq:Mc} is not symmetric necessitates the consideration of the real and imaginary parts of the Fourier-transformed fluctuation amplitudes. Regrouping the fluctuation amplitudes in the vector $\Psi_{cq}=(\Psi_{cq}^\prime,\Psi_{cq}'')^{\top}$ where $\Psi_{cq}^\prime= (d_{1q}^\prime,d_{2q}'',p_{0q}^\prime,\beta_{0q}^\prime)$ and $\Psi_{cq}''= (d_{1q}'',d_{2q}^\prime,p_{0q}'',\beta_{0q}'')$, the $8\times 8$ fluctuation matrix decomposes into two $4\times 4$ blocks $\mathcal{M}_{c}^\prime(q)$ and $\mathcal{M}_{c}''(q)=[\mathcal{M}_{c}'(q)]^{\ast}$
\begin{equation}
\mathcal{M}_{c}^\prime(q)=\left(
\begin{array}
[c]{cccc}%
M_{11} & iM_{12} & M_{13} & -iM_{14}\\
iM_{12} & M_{22} & -iM_{23} & -iM_{24}\\
M_{13} & -iM_{23} & M_{33} & M_{34}\\
-iM_{14} & -iM_{24} & M_{34} & M_{44}
\end{array}
\right),
\end{equation}
where $M_{\mu\nu}$ are the elements of the matrix $\mathcal{M}_{c}$ defined in \eqref{eq:Mc}. The four real-valued amplitudes entering $\Psi_{cq}^\prime$ are integrated from $-\infty$ to $\infty$. The result of the functional integration for the generating functional is therefore
\begin{align}
Y_{fl,cont}^{(c)}(J_{c,\mu})  & =c\left(  \prod_{q}Y_{q}^{(c)}%
(J_{c,\mu})\right)  ^{1/2},\\
Y_{q}^{(c)}(J_{c,\mu})  & =\frac{1}{|\det{M_{c}(q)^\prime}|}\exp\{J_{c,\mu}^\prime M_{c,\mu,\nu}^{\prime -1}J_{c,\nu}^\prime\ + J_{c,\mu}^{\prime\prime} M_{c,\mu,\nu}^{\prime\prime -1}J_{c,\nu}^{\prime\prime}\},\nonumber
\end{align}
where $M_{c,\mu,\nu}^{\prime -1}(q)$ are the elements of the matrix inverse  and $c$ is a constant. 
We note that $\det{\mathcal{M}_{c}^\prime(q)}=\det{\mathcal{M}_{c}(q)}=\det{\mathcal{M}_{c}''(q)}$.
The charge response function is obtained as 
\begin{align}
	\chi_c(q) &= 8\mf d^2 \chi_{d_{1}d_{1}}(q) +2\mf p_0^2 \chi_{p_{0}p_{0}}(q) + 4\mf d_{\phantom{0}}\mf p_0\chi_{d_{1}p_{0}}(q).
\end{align}

We note in passing that the two-particle correlation functions $\chi_{s}(q), \chi_{c}(q)$ are found to adopt the correct limiting behavior $\chi_{0}(q)$ in the limit $U=0$. The four-particle correlation functions $\chi_{d_{1}d_{1}}(q), \chi_{p_{0}p_{0}}(q)$ do not reduce to the correct noninteracting limit. This limit would be given by $\lim_{U\rightarrow 0}\chi_{d_{1}d_{1}}(q) \propto \int_{q_{1}} \chi_{0}(q-q_{1})\chi_{0}(q_{1})$, an expression that is beyond the reach of the Gaussian approximation.

\section{High frequency contributions}\label{app:sec:HFC}

The HFC arise from the bosonic and the pseudo-fermionic part of the Hamiltonian. We now consider the contributions from the spin and charge sectors separately.

\subsection{HFC in the spin sector}
The bosonic part of the HFC to the grand potential per site counting the three
spin directions is given by 
\begin{equation}\label{app:eq:HFCsb}
\Omega^{(2)}_{HFC,3s,b}=3(\beta _{0}-\alpha )/2.
\end{equation}
The pseudo-fermionic terms involve expansion of the $z$-factors with respect to fluctuations. The fluctuation terms are expressed as 
\begin{align}
\underline{L}& =L_{0}\left[\underline{\tau }_{0}+\frac{1}{4}\frac{(\mathbf{p}%
^{\dag }\mathbf{p)}\underline{\tau }_{0}+((\mathbf{p\underline{\mathbf{\tau }%
})+(p}^{\dag }\underline{\mathbf{\tau }}))\overline{p}_{0}}{1-\overline{d}%
^{2}-\overline{p}_{0}^{2}/2}\right], \\
\underline{R}& =R_{0}\left[\underline{\tau }_{0}+\frac{1}{4}\frac{(\mathbf{p}%
^{\dag }\mathbf{p)}\underline{\tau }_{0}-((\mathbf{p\underline{\mathbf{\tau }%
})+(p}^{\dag }\underline{\mathbf{\tau }}))\overline{p}_{0}}{1-\overline{e}%
^{2}-\overline{p}_{0}^{2}/2}\right].
\end{align}%
The terms linear in $\mathbf{p}^{\dag },\mathbf{p}$ are relevant because
they get multiplied by the linear terms of the external operators $%
\underline{p},$ $\widetilde{\underline{p}}$ . These arise as 
\begin{align}
((\mathbf{p\underline{\mathbf{\tau }})+(p}^{\dag }\underline{\mathbf{\tau }}%
))\overline{p}_{0}\underline{p}& =\frac{\overline{p}_{0}}{2}\left[(\mathbf{p}%
^{\dag }\underline{\mathbf{\tau }})(\mathbf{p}\underline{\mathbf{\tau }}%
)+...\right], \\
((\mathbf{p\underline{\mathbf{\tau }})+(p}^{\dag }\underline{\mathbf{\tau }}%
))\overline{p}_{0}\widetilde{\underline{p}}^{\dag }& =-\frac{\overline{p}_{0}%
}{2}\left[(\mathbf{p\underline{\mathbf{\tau }})}(\mathbf{p}^{\dag }\underline{%
\mathbf{\tau }})+...\right], \\
\widetilde{\underline{p}}^{\dag }((\mathbf{p}\underline{\mathbf{\tau }})+(p%
^{\dag }\underline{\mathbf{\tau }}))\overline{p}_{0}& =-\frac{\overline{p}%
_{0}}{2}\left[(\mathbf{p}^{\dag }\underline{\mathbf{\tau }})(\mathbf{p}\underline{%
\mathbf{\tau }})+...\right], \\
\overline{p}_{0}\underline{p}((\mathbf{p\underline{\mathbf{\tau }})+(p}%
^{\dag }\underline{\mathbf{\tau }}))& =\frac{\overline{p}_{0}}{2}\left[(\mathbf{p%
\underline{\mathbf{\tau }})}(\mathbf{p}^{\dag }\underline{\mathbf{\tau }}%
)+...\right].
\end{align}%
We note that both orderings, $(\mathbf{p}^{\dag }\underline{\mathbf{\tau }})(%
\mathbf{p\underline{\mathbf{\tau }})}$ and $(\mathbf{p\underline{\mathbf{%
\tau }})}(\mathbf{p}^{\dag }\underline{\mathbf{\tau }})$ appear.  In
principle, second-order derivative terms might also contribute. They drop
out, however, because they lead to symmetric operator product combinations
which do not have HFCs, for example 
\begin{equation*}
((\mathbf{p\underline{\mathbf{\tau }})+(p}^{\dag }\underline{\mathbf{\tau }}%
))((\mathbf{p\underline{\mathbf{\tau }})+(p}^{\dag }%
\underline{\mathbf{\tau }}))=\left[(\mathbf{p}%
^{\dag }\underline{\mathbf{\tau }})(\mathbf{p\underline{\mathbf{\tau }})+}(%
\mathbf{p\underline{\mathbf{\tau }})}(\mathbf{p}^{\dag }\underline{\mathbf{%
\tau }})\mathbf{+...}\right].
\end{equation*}%
The HFC of the $z$-factor is then obtained by replacing $p_{\mu }^{\dag }p_{\mu
}\rightarrow -1/2$, $p_{\mu }p_{\mu }^{\dag }\rightarrow +1/2$\ such that $(%
\mathbf{p}^{\dag }\mathbf{p)}\underline{\tau }_{0}-(\mathbf{p\underline{%
\mathbf{\tau }})}(\mathbf{p}^{\dag }\underline{\mathbf{\tau }})\mathbf{%
\rightarrow }-3\underline{\tau }_{0}$ and $(\mathbf{p}^{\dag }\mathbf{p)}%
\underline{\tau }_{0}+(\mathbf{p}^{\dag }\underline{\mathbf{\tau }})(\mathbf{%
p\underline{\mathbf{\tau }})\rightarrow -}3\underline{\tau }_{0}$: 
\begin{equation}
z_{HFC,3s,f}=-\frac{3\overline{z}}{4}\left[\frac{1}{1-\overline{d}^{2}-\overline{p}%
_{\sigma }^{2}}+\frac{1}{1-\overline{e}^{2}-\overline{p}_{-\sigma }^{2}}\right].
\end{equation}%
The contribution to the grand potential from the pseudo-fermionic sector is found as 
\begin{equation}
\Omega^{(2)}_{HFC,3s,f}=-2z_{HFC,3s,f}\overline{z}\varepsilon_0,
\end{equation}
where the characteristic pseudo-fermionic energy is defined as $\varepsilon_0 = -\frac{2}{N}\sum_{\bk} n_F(\epsilon_\bk)\xi_\bk$ and $\overline{z}$ is the mean field value of $z$.

\subsection{HFC in the charge sector}
The bosonic part of the HFC to the grand potential per site counting the three bosons $d,e,p_{0}$ is given by 
\begin{equation}\label{app:eq:HFCcb}
\Omega^{(2)}_{HFC,c,b}=-U/2+3(\beta _{0}-\alpha )/2.
\end{equation}%
The terms linear in $d^{\dag },d,e,p_{0}$ are relevant because they get multiplied
by the linear terms of the external operators $d^{\dag },d,e,p_{0}$. These arise as
\begin{align}
\begin{split}
\overline{p}^\nodag_{0}(p_{0}^{\dag }+p_{0}^\nodag)p_{0}^\nodag& =\overline{p}_{0}^\nodag p_{0}^{\dag
}p_{0}^\nodag+..., \\
\overline{p}_{0}^\nodag p^\nodag_{0}(p_{0}^{\dag }+p^\nodag_{0})& =\overline{p}^\nodag_{0}p^\nodag_{0}p_{0}^{%
\dag }+... \\
e^{\dag }\overline{e}(e^{\dag }+e)& =\overline{e}e^{\dag }e+..,   \\
\overline{d}(d^{\dag }+d)d& =\overline{d}d^{\dag }d+...  
\end{split}
\end{align}
In principle, second-order derivative terms might also contribute. They drop out, however, because they lead to symmetric operator product combinations which do not have HFCs. 
Substituting these results into the expression for $\underline{z}$, keeping only the terms $p_{0}^{\dag }p^\nodag_{0}$ and $p^\nodag_{0}p_{0}^{\dag }$ , etc, and dropping higher order terms such as $(p_{0}^{\dag }p^\nodag_{0})^{2}$ in the present Gaussian approximation, we get 
\begin{align}
\begin{split}
\sqrt{2}\underline{L}e^{\dag }\underline{p}\underline{R}
=\underline{\tau }_{0}\frac{\overline{e}\overline{z}}{(\overline{e}+\overline{d})}\Bigg{[}1
&+\frac{1}{2}\frac{p_{0}^{\dag }p^\nodag_{0}\mathbf{+}d^{\dag }d}{(1-\overline{d}^{2}-\overline{p}_{0}^{2}/2)} \\ 
&+\frac{1}{2}\frac{\frac{1}{2}(p_{0}^{\dag
}p^\nodag_{0}+p_{0}p_{0}^{\dag })+2e^{\dag }e}{(1-\overline{e}^{2}-\overline{p}%
_{0}^{2}/2)}\Bigg{]}.
\end{split}
\end{align}
Likewise 
\begin{align}
\begin{split}
\sqrt{2}\underline{L}\widetilde{\underline{p}}^{\dag }d\underline{R}
=\underline{\tau }_{0}\frac{\overline{d}\overline{z}}{(\overline{e}+\overline{d})}\Bigg{[}1
&+\frac{1}{2}\frac{p_{0}^{\dag }p^\nodag_{0}+p^\nodag_{0}p_{0}^{\dag }+2d^{\dag }d}{(1-\overline{d}^{2}-\overline{p}_{0}^{2}/2)} \\
&+\frac{1}{2}\frac{p_{0}^{\dag }p^\nodag_{0}\mathbf{+}e^{\dag }e}{(1-\overline{e}^{2}-\overline{p}_{0}^{2}/2)}\Bigg{]}.
\end{split}
\end{align}
The HFC of $z$ is then obtained, replacing $p_{0}^{\dag }p_{0}^\nodag,e^{\dag}e,d^{\dag }d\rightarrow -1/2$ and $p_{0}^\nodag p_{0}^{\dag }\rightarrow +1/2$: 
\begin{equation}
z_{HFC,c,f} =-\frac{\overline{z}}{2}\left[\frac{1}{1-\overline{d}^{2}-\overline{p}_{0}^{2}/2}+%
\frac{1}{1-\overline{e}^{2}-\overline{p}_{0}^{2}/2}\right]. 
\end{equation}
The contribution to the grand potential from the pseudo-fermionic charge sector is found as 
\begin{equation}
\Omega^{(2)}_{HFC,c,f}=-2z_{HFC,c,f}\overline{z}\varepsilon_0.
\end{equation}
In total, the HFC to the grand potential may be summed up to give 
\begin{equation}
\Omega^{(2)}_{HFC}=-U/2+3(\beta _{0}-\alpha )+\Omega^{(2)}_{HFC,3s,f}+\Omega^{(2)}_{HFC,c,f}.
\end{equation}
The correction turns out to be substantial, of the order of the mean field $\Omega^{(0)}$.

\section{Free bosons}\label{app:sec:freebosons}

In this section, we display the working of the evaluation of the functional integral representation of the partition function in terms of CT approach plus HFC in a most simple example, a free boson system. 
The Hamiltonian for free bosons in quantum states labeled by $\alpha$ is given by
\begin{equation}
	\hat{H} = \sum_\alpha \epsilon_\alpha \, b_\alpha^\dagger b_\alpha^\nodag.
\end{equation}
In the following, we determine the grand potential of this system from its functional integral representation in three different ways and recover the same, well-known result. The first two approaches are textbook-knowledge, i.e., (i) the rigorous calculation in discrete imaginary time and (ii) complex integration in the CT limit with an adequate convergence parameter to cure the ill-defined limit to continuous time. Both methods are hard to generalize to the numerical evaluation of non-analytically solvable models. Lastly, we (iii) evaluate the grand potential in the CT limit without a convergence parameter and recover the correct result with the addition of high frequency contributions (HFC). This approach is in close analogy to our approach of calculating fluctuation corrections to the grand potential of the Hubbard model described in this paper.

\subsection{Discrete imaginary time}

Within the exact discrete imaginary-time path integral, the partition function is given by \eqref{eq:Zdiscrete} with the action in \eqref{eq:Sdiscrete}. After integration of the fields, the partition function is found to be the inverse action determinant
\begin{equation}
Z = \prod_\alpha \lim_{M\rightarrow \infty} \left(\det  S \right)_\alpha^{-1},
\end{equation}
where the $S$ is represented by the $M\times M$ matrix
 \begin{subequations}
 	\begin{align}
 	S
 	&=
 	\begin{pmatrix}
 	1 & 0 &0& \cdots & 0 & - \gamma\\
 	-\gamma & 1 &0&   &  & 0 \\
 	0 &-\gamma &1&\ddots & & \vdots\\
 	\vdots& 0 & -\gamma&\ddots  &  \\
 	& & 0& \ddots & & 0\\ 
 	0& & &    & -\gamma &1 \\
 	\end{pmatrix},
 	\intertext{
 		with 
 	}
 	\gamma&= 1- \frac{\beta}{M}\epsilon_\alpha,
 	\intertext{and $M$ is the number of time slices. By Laplace expansion of the first row, the determinant is found to be
 	}
 	\det S&=1+(-1)^{M-1}(-\gamma)^M .
 	\intertext{Evaluating the limit of infinite time slices, the partition function of free particles is found to be
 	}
 	\label{SBT:eq:Z0}
 	Z&=\prod_\alpha \left(1- \e^{-\beta \epsilon_\alpha} \right)^{-1},
    \end{align}
 \end{subequations}
yielding the well-known result
\begin{equation}
    \Omega = - T \ln Z =  T \sum_{\alpha} \ln\left(1- \e^{-\beta \epsilon_\alpha} \right).
\end{equation}

\subsection{Continuous imaginary time with convergence factor}\label{sec:FreebosonCT}

In the CT limit, the action is diagonalized by Fourier transformation to Matsubara space and given by
\begin{equation}
    S = T\sum_\alpha \sum_{\i\omega_n} b^*_{n,\alpha} (-\i\omega_n + \epsilon_\alpha) b_{n,\alpha}.
\end{equation}
By integrating out the fields $b_{n,\alpha}$, we find the partition function 
\begin{equation}
    Z = \prod_{\alpha}\prod_{\omega_n} \frac{1}{-\i\omega_n + \epsilon_\alpha},
\end{equation}
which is formally divergent due to the ill-defined continuum limit. The correct result for the grand potential can be recovered by mapping onto a complex contour integral and the addition of the convergence factor $\e^{\omega_n 0^+}$:
\begin{align}\label{SBT:eq:integratef}
    \begin{split}
	\Omega&= T \sum_{\alpha,n}  \ln(-i\omega_n+\epsilon_\alpha)\e^{i\omega_n 0^+}\\
	&=
	- \sum_\alpha \oint \frac{dz}{2\pi i} n_{B}(z) \ln\left(\epsilon_\alpha-z\right) \e^{z0^+}\\
	&= - \sum_\alpha \int_{-\infty}^\infty \frac{d\omega}{2\pi \i} n_{B}(\omega) \left[\ln\left(\epsilon_\alpha -\omega +i0^+ \right) -\ln\left(\epsilon_\alpha -\omega -i0^+ \right) \right]\\
	&=
	- \sum_\alpha \int_{-\infty}^\infty \frac{d\omega}{2\pi \i} n_{B}(\omega) 2\pi\i\ \Theta\left(\omega-\epsilon_\alpha \right)
	= -\sum_\alpha \int_{\epsilon_\alpha}^\infty d\omega n_{B}(\omega)\\
	&= T \sum_\alpha \ln \left(1 - \e^{-\beta \epsilon_\alpha} \right).
	\end{split}
\end{align}

\subsection{Continuous imaginary time with high frequency contributions}\label{app:sec:CTHFC}

In close analogy to the $d_1,d_2$ fields in the SB formalism, we now decompose the bosonic operators for the free boson Hamiltonian into their real and imaginary part $b=b_1+\i b_2$. This way, the action can be represented by a $2\times 2$ matrix
\begin{equation}
	\mathcal{S} =\sum_\alpha \sum_{\i\omega_n}
	\begin{pmatrix}
	b_1^*& b_2^*
	\end{pmatrix}
	\begin{pmatrix}
\epsilon_\alpha & \omega_n \\
-\omega_n & \epsilon_\alpha
\end{pmatrix}
	\begin{pmatrix}
b_1 \\ b_2
\end{pmatrix},
\end{equation}
whose determinant
\begin{equation}
    D(\omega_n) = \epsilon_\alpha^2 + \omega_n^2 = (\epsilon_\alpha +\i\omega_n)(\epsilon_\alpha -\i\omega_n)
\end{equation}
is an even function in $\omega_n$ like the fluctuation determinant $\det\mathcal{M}$. The partition function is given by
\begin{equation}
    Z = \prod_{\alpha}\prod_{\omega_n} \frac{1}{\sqrt{D(\omega_n)}},
\end{equation}
the square root is added because the determinant $D(\omega)$ is already a product of positive and negative Matsubara frequencies such that there would be an over-counting otherwise in this basis.
The grand potential is thus given by
\begin{equation}
\Omega_{CT} =T\sum_\alpha  \frac12\sum_{i\omega_n} \ln D(\omega_n).
\end{equation}
In full analogy to the procedure discussed in \autoref{sec:CT}, the Matsubara sum is mapped onto a complex contour integral. After integration by parts, one finds
\begin{equation}
\Omega = \frac12 \sum_\alpha \int_{-\infty}^\infty \frac{d\omega}{\pi} f(\omega) S(\omega),
\end{equation}
where
\begin{align}
f(\omega) &= \frac{\omega}{2}+ T\ln \Big{|} 1-\e^{-\omega/T} \Big{|},\\
S(\bq,\omega) &= \frac{-D'\partial_\omega D'' + D'' \partial_\omega D'}{|D|^2},
\end{align}
with $D= D'+ i D''$ within the analytic continuation $\i\omega_n \rightarrow \omega + i \eta$ and $\eta\rightarrow 0^+$. For the determinant of free bosons, the distribution of bosonic excitations is given by
\begin{align}
	S(\omega) &=\lim_{\eta\rightarrow 0} \frac{\eta}{\eta^2+(\omega_n-\epsilon_\alpha^2)}+\frac{\eta}{\eta^2+(\omega_n+\epsilon_\alpha^2)}\\
 &= 
 \pi \delta(\omega-\epsilon_\alpha) + \pi\delta(\omega+\epsilon_\alpha).
\end{align}
Since the integrand is a symmetric function in $\omega$, we can rewrite the integral boundaries
\begin{equation}
	\Omega_{CT} = \sum_\alpha  \int_{0}^\infty \frac{d\omega}{\pi} f(\omega) S(\omega)
	= \sum_\alpha \frac{\epsilon_\alpha}{2}+  T \sum_\alpha \ln \left(1 - \e^{-\beta \epsilon_\alpha}\right).
	\end{equation}
Compared to the previous approaches, we have the additional zero-point energy $\sum_\alpha \frac{\epsilon_\alpha}{2}$, which is not present in the exact discrete-time result. This is due to the ill-defined continuum limit. The high frequency contributions (HFC) described in \autoref{sec:HFC} are given by the commutator
\begin{equation}
\Omega_{HFC}=\frac12	\sum_\alpha \epsilon_\alpha [b^\dagger,b] =-\frac12	\sum_\alpha \epsilon_\alpha.
\end{equation}
Therefore, we recover the correct result
\begin{equation}
\Omega = \Omega_{CT} + \Omega_{HFC} =  
T \sum_\alpha \ln \left(1 - \e^{-\beta \epsilon_\alpha}\right).
\end{equation}
This demonstrates the effectiveness of the HFC by correctly complementing the CT result in an analytically controlled example. 

\section{Non-interacting limit ($U=0$)} \label{sec:app:U0}

 Here, we show that the non-interacting limit $(U=0)$, where the analytic solution of the Hubbard model is available, is fully recovered within the presented KR slave-boson mean field and fluctuation formalism (SRIKR+). Specifically, in this limit, (i) the MF saddle point recovers the exact solution, (ii) the fluctuation corrections to the ground-state energy vanish, and (iii) the dynamic spin $\chi_s(q)$ and charge susceptibility $\chi_c(q)$ reduce to the bare susceptibility $\chi_0(q)$, which is the expected behavior.

For free electrons that obey the Pauli principle, the states $\ket{\uparrow}$, $\ket{\downarrow}$ are occupied randomly with the same probability on each lattice site. For an average filling of $n$ per site, where $ 0 \leq n \leq 2$, the mean occupation of both states is given by $n/2$, respectively. Consequently, the average number of doubly occupied states is given by $d^2=(n/2)^2$. The mean value of the density of singly occupied sites yields $p_0^2 = 2(n/2)(1-n/2)$ by accounting for the spin degeneracy. Finally, we find $e^2 = 1- d^2 - p_0^2 = (1 - n/2)^2$. The MF values derived from this simple argument are consistent with the analytic solution presented in \eqref{eq:MFsolution}. Moreover, the Lagrange multiplier $\beta_0$ vanishes at $U=0$, whereas $\alpha$ remains finite and is a positive definite quantity. In summary, the density-dependent MF solution in the non-interacting limit is given by
\begin{align}\label{eq:MFsolutionU0}
\begin{split}
 e &= 1-\frac{n}{2},\, p_0 = \sqrt{n\left(1-\frac{n}{2}\right)}, \,  d = \frac{n}{2}, \, \bs p =0,  \\
 \beta_0 &= 0, \,\bs\beta = 0,\\
  \alpha &=  \frac{4}{n(2-n)}\varepsilon_0, 
  \end{split}
\end{align}
where $\varepsilon_0$ has been defined in \eqref{eq:epsilon0}. Within the MF ansatz, the Kotliar-Ruckenstein renormalized hopping amplitude defined in \eqref{eq:KRzfactor} reduces to 
\begin{equation}
z_0 = \frac{p_0(e+d)}{\sqrt{2(1-d^2-p_0^2/2)(1-e^2-p_0^2/2)}} = 1\Bigg{|}_{U=0}
\end{equation}
and is equal to unity employing the values for the boson amplitudes given in \eqref{eq:MFsolutionU0} (equivalent to the values cited above) for arbitrary filling $n$. Thus, the non-interacting limit is exactly fulfilled on MF level. For fluctuations around PM saddle points, the spin and charge sectors are decoupled and independently satisfy the exact non-interacting limit, as we show in the following.

\subsection{Spin sector}

With \eqref{eq:MFsolutionU0}, we can evaluate all $z$-factor derivatives that appear in the fluctuation calculation such that the results depend solely on $n$. We find $\tilde{z}_{+,p_1p_1}=\tilde{z}_{-, p_1}=0$ for arbitrary filling such that the spin fluctuation matrix given by \eqref{eq:Ms}
simplifies to
\begin{equation}\label{eq:MsU0}
    \mathcal{M}_s(q)\Big{|}_{U=0} = 
    \begin{pmatrix}
    0 & -\mf p_0 \\
    -\mf p_0 & -\chi_0(q)/2
    \end{pmatrix}.
\end{equation}
Consequently, the spin susceptibility
\begin{equation}
    \chi_s(q) =2 \mf p_0^2\mathcal{M}^{-1}_{p_1p_1} = \chi_0(q)\Big{|}_{U=0},
\end{equation}
reduces to the bare susceptibility $\chi_0(q)$ as it is expected in the non-interacting limit. 

\subsubsection{Continuous time contribution}

The spin determinant 
\begin{equation}
  \det \mathcal{M}_s\Big{|}_{U=0} = -p_0^2  
\end{equation}
does not depend on $\omega_n$ such that the CT contribution to the grand potential given by \eqref{eq:OmegaCT} vanishes
\begin{equation}
  \Omega^{(2)}_{CT,s}\Big{|}_{U=0}= 0.
\end{equation}
Notice that the sign of the determinant depends on the integration path of the Lagrange multipliers. The results given here correspond to integration along the real axis, while the Lagrange multipliers $\beta_0$ and $\bs \beta$ actually need to be integrated through the saddle point along the imaginary axis. This changes the fluctuation matrix according to $\mathcal{M}_{\mu,\beta_\nu}\rightarrow \i\mathcal{M}_{\mu,\beta_\nu},\mathcal{M}_{\beta,\mu_\nu}\rightarrow \i\mathcal{M}_{\beta,\mu_\nu}$ and $\mathcal{M}_{\beta_\nu,\beta_\nu}\rightarrow -\mathcal{M}_{\beta_\nu,\beta_\nu}$, where $\nu=0,1,2,3$. With that applied, the spin determinant changes its sign and is positive definite as expected for stable saddle points. Since the susceptibilities and fluctuation corrections are independent of this transformation, we kept the fluctuation matrix in the convention that was applied in previous works.

\subsubsection{High frequency contribution}

According to \eqref{eq:OmegaHFCs}, the HFC contribution 
\begin{align}
\Omega^{(2)}_{HFC,s} &= -\frac{3}{2}\left(\alpha-\beta_0\right) -2 z_{HFC,3s}z_0 \varepsilon_0\\
&=-\frac{3}{2}\alpha +  \frac{6}{n(2-n)} \varepsilon_0 \Big{|}_{U=0}= 0
\end{align}
vanishes for any filling $n$ in the non-interacting limit, which can be verified by inserting \eqref{eq:MsU0} into the expression for $z_{HFC,3s}$ given by \eqref{eq:ZHFCs}. Thus, there are no spin fluctuation contributions at $U=0$.

\subsection{Charge sector}

 For the intricate charge fluctuation matrix derived in \app{app:sec:fluctuationmatrix}, there are extensive simplifications at $U=0$ because the first derivatives vanish $\tilde{z}_{d_1}=\tilde{z}_{p_0}=0$. This can be directly inferred from the MF saddle point equations $\partial \Omega^{(0)}/\partial \psi_\mu =0 $, where $\alpha$ does not enter because of the substitution of the $e$-field and $\beta_0=0$ such that only the pseudofermionic part proportional to the derivative remains. Notice that $z_{d_1}\neq 0$, $z_{p_0}\neq 0$ because $\alpha$ enters the saddle point equations without the substitution of the $e$-field (or any other field). The derivatives wrt.~$d_2$ do not vanish but are purely imaginary and yield $\tilde{z}_{d_2} =i$, $\tilde{z}^*_{d_2} =-i$ for any $n$. The second derivatives are density dependent and can be evaluated with \eqref{eq:MFsolutionU0}. Consequently, the charge fluctuation matrix given by \eqref{eq:Mc} is simplified to
\begin{widetext}
\begin{equation}\label{eq:McU0}
\mathcal{M}_c(q)\Big{|}_{U=0} 
=
\begin{pmatrix}
-\varepsilon_0 \tilde{z}_{d_1d_1} &\omega_n  & -\varepsilon_0 \tilde{z}_{d_1p_0} &-n \\
-\omega_n	& -\varepsilon_0(1+\tilde{z}_{d_2d_2})+ \frac12 \omega_n^2 \,\chi_0(q)& 0 & -\frac12\omega_n\,\chi_0 (q)\\
-\varepsilon_0 \tilde{z}_{d_1d_1} & 0 & -\varepsilon_0 \tilde{z}_{p_0p_0} &-\sqrt{n(1-\frac{n}{2})} \\
-n & \frac12 \omega_n \,\chi_0 & -\sqrt{n(1-\frac{n}{2})} & -\frac12\chi_0(q)
\end{pmatrix},
\end{equation}
\end{widetext}
 where
 \begingroup
\allowdisplaybreaks
 \begin{align}
\tilde{z}_{d_1d_1} &= -\frac{16 ( 1-n )^2}{(2 - n)^3 n}, \\
\tilde{z}_{d_1p_0}& =\frac{16 (1 - n)}{(2 - n) \sqrt{(2 - n)^3 2 n}},\\
\tilde{z}_{d_2d_2} &= - \frac{2}{n},\\
\tilde{z}_{p_0d_0} &= - \frac{8}{(2-n)^2}.
 \end{align}
 \endgroup
The charge susceptibility is given by
\begin{align}
	\chi_c(q) &= 8\mf d^2 \mathcal{M}_{d_1,d_1}^{-1}(q) +2\mf p_0^2 \mathcal{M}_{p_0,p_0}^{-1}(q) + 4\mf d_{\phantom{0}}\mf p_0\mathcal{M}_{d_1,p_0}^{-1}(q)\nonumber\\
 &= \chi_0(q)\Big{|}_{U=0},
\end{align}
and thus reduces to the bare susceptibility in analogy to the spin sector. We confirmed that this result is basis-independent; it is also recovered within the basis of complex fields $d^*,d$ rather than the decomposition in real and imaginary part $d_1,d_2$ that we apply. 

\subsubsection{Continuous time contribution}

The charge determinant inferred from \eqref{eq:McU0}
\begin{align}
\begin{split}
\det\mathcal{M}_c(\omega_n)\Big{|}_{U=0} 
&= \frac{n(n-2)}{2} \left[\left( \frac{4\epsilon_0}{n(n-2)}\right)^2 +\omega_n^2\right] \\
&= - p_0^2 \left(\alpha^2+\omega_n^2 \right)
\end{split}
\end{align}
does not depend on the momentum $\bq$ since $\chi_0(q)$ does not enter the equation. Since the prefactor drops out within the analytical continuation, this determinant exactly corresponds to the case of free bosons discussed in \app{app:sec:CTHFC}, whereby the bosonic energy is given by $\alpha(n)$. Consequently, the CT contribution to the grand potential is given by
\begin{equation}
    \Omega^{(2)}_{CT,c}\Big{|}_{U=0} =  \frac{\alpha}{2} + T\log\left(1-\e^{-\alpha/T}\right)\rightarrow \frac{\alpha}{2}.
\end{equation}
The last equation holds in the limit of low temperatures, which we only consider here. Finite temperature corrections are exponentially small at $T\ll \alpha$. We have verified that this result is basis independent: Applying the $d,d^*$ basis changes the determinant only by a constant factor that drops out in the analytic continuation. Moreover, the determinant of the $6\times 6$ charge matrix that adds $(e,\alpha)$ to the fluctuation basis also differs only by a constant factor. At $U=0$: $\det\mathcal{M}_c^{6\times 6}= \mf e^2 \det\mathcal{M}_c^{4\times 4}$, such that the CT fluctuation contribution is also independent of that basis choice.

\subsubsection{High frequency contribution}

According to \eqref{eq:OmegaHFCc}, the HFC for the charge sector is given by
\begin{align}
\begin{split}
\Omega^{(2)}_{HFC,c} &= -\frac{U}{2}-\frac{3}{2}\left(\alpha-\beta_0\right) -2 z_{HFC,c}z_0 \varepsilon_0\\
&=-\frac{3}{2}\alpha +  \frac{4}{n(2-n)} \varepsilon_0\Bigg{|}_{U=0} 
= -\frac{\alpha}{2},
\end{split}
\end{align} which can be verified by inserting the non-interacting MF solution into the expression for $z_{HFC,c}$ given by \eqref{eq:ZHFCc}. In contrast to the spin sector, the CT and HFC contributions do not vanish independently, but their sum
\begin{equation}
\Omega^{(2)}_{c}\Big{|}_{U=0} =  \Omega^{(2)}_{CT,c}\Big{|}_{U=0}+\Omega^{(2)}_{HFC,c}\Big{|}_{U=0}=0
\end{equation}
is zero, such that there are no charge fluctuation contributions at $U=0$. We note that the above cancellation does not hold any more at temperatures of the order of $\alpha$. At such high temperatures the Gaussian approximation breaks down and the interaction of fluctuations becomes important.

\section{Fluctuation corrections with nearest-neighbor interaction $V$} \label{app:sec:V}

We calculate the CT and HFC fluctuation contributions wrt.~the nearest-neighbor density-density interaction
\begin{equation}\label{app:eq:HV}
    H^V = V \sum_{<ij>} \hat{n}_i\hat{n}_j,
\end{equation}
where $\hat{n}_i= \sum_\sigma f^\dagger_{i,\sigma }f^\nodag_{i,\sigma } = 2d_i^2+p_{0,i}^2+\bs p^2_i$. As shown in Ref.~\cite{Riegler_2023}, the MF solution for the bosons is invariant under this interaction. Only the chemical potential and the constraint parameter $\beta_{0}$ change according to
\begin{align}\label{app:eq:beta0V}
    \beta_0 &= \beta_0\Big{|}_{V=0}+4Vn,\\
    \mu_0 &= \mu_0\Big{|}_{V=0}+4Vn.
\end{align}
For the bosonic representation of the charge density $\hat{n}_i=2d_i^2+p_{0,i}^2+\bs p^2_i$ that must be considered in the HFC, the MF solution of the Lagrange multiplier $\alpha$ remains invariant.

\subsection{High frequency contribution}

As we show in the following, the HFC is independent of $V$. We expand \eqref{app:eq:HV} in anlogy to the previous HFC derivation
\begin{equation}\label{app:eq:HFCHV}
H^V \rightarrow 4 V n (2d^\dagger d + p_0^\dagger p^\nodag_0 + \bs p^\dagger \bs p).   
\end{equation}
The HFC from this term is inferred by replacing the number operators by, e.g., $d^\dagger d \rightarrow -\frac12$.

\subsubsection{Spin sector}

The HFC from \eqref{app:eq:HFCHV} for the spin sector with the fields $p_1,p_2,p_3$ is given by $-\frac32 4Vn$. With the additional shift of $\beta_0$ given by \eqref{app:eq:beta0V} 
and the solution at $V=0$ given by \eqref{app:eq:HFCsb}, we find
\begin{equation}
    \Omega^{(2)}_{HFC,s}(V)=\Omega^{(2)}_{HFC,s}(V=0),
\end{equation}
i.e., independent of $V$. Notice that $z_{HFC,3s}$ is independent of $V$ since the MF value of $e,d,p_0,\bs p$ does not depend on $V$.

\subsubsection{Charge sector}

The HFC from \eqref{app:eq:HFCHV} for the charge sector with the fields $d,e,p_0$ is given by $-\frac32 4Vn$. With the additional shift of $\beta_0$ given by \eqref{app:eq:beta0V} 
and the solution at $V=0$ given by \eqref{app:eq:HFCcb}, we again find
\begin{equation}
    \Omega^{(2)}_{HFC,c}(V)=\Omega^{(2)}_{HFC,c}(V=0).
\end{equation}
Notice that $z_{HFC,c}$ is independent of $V$ since the MF value of $e,d,p_0,\bs p$ does not depend on $V$.

\subsection{Continuous time contribution}

As we show below, the CT contribution of the charge sector depends on $V$.

\subsubsection{Spin sector}

In Ref.~\cite{Riegler_2023} it has been shown that the (transversal) spin channel of the CT fluctuation matrix does not depend on $V$. Consequently $\Omega_{CT,s}$ is independent of $V$.

\subsubsection{Charge sector}

As shown in Ref.~\cite{Riegler_2023}, the CT fluctuation matrix employs additional frequency-independent matrix elements in the charge sector that are proportional to $V$. In the basis $(d_1,d_2,p_0,\beta_0)$ applied previously, these are given by
\begin{equation}\label{app:eq:McV}
\mathcal{M}_{c}^V(\bq)
=V
\begin{pmatrix}
8d^2 f(\bq) &0  & 4dp_0f(\bq)&0  \\
0	& 0 & 0 &0\\
4dp_0f(\bq) & 0 & 2p_0^2f(\bq) &0 \\
0 & 0 & 0 & 0  \\
\end{pmatrix} ,
\end{equation}
where 
\begin{equation}
    f(\bq) = 2(\cos q_x  + \cos q_y)
\end{equation}
is the form-factor for the nearest-neighbor interaction that can be generalized to longer-ranged interactions \cite{Riegler_2023}. \eqref{app:eq:McV} already accounts for the shift of $\beta_0$ such that the total charge fluctuation matrix for $U,V$ is given by
\begin{equation}
    \mathcal{M}_c^{U,V} = \mathcal{M}_c(q)\Big{|}_{V=0} + \mathcal{M}^V_c(\bq),
\end{equation}
where $\mathcal{M}_c(q)$ is given by \eqref{eq:Mc}. At $U=0$, the total charge determinant is given by
\begin{equation}
 \det\mathcal{M}_c^{U,V}\big{|}_{U=0} = -p_0^2 \left[ 1+ V f(\bq)\chi_0(q)\right]\left[\omega_n^2+\alpha^2\right].   
\end{equation}
The free particle contribution $\alpha^2+\omega_n^2$ can be separated and cancels with the HFC such that
\begin{equation}
    \Omega^{(2)}(V)\big{|}_{U=0} =  \frac1N\sum_\bq \frac12 T\sum_{i\omega_n} \ln\left[1+Vf(\bq)\chi_0(\bq,i\omega_n) \right].
\end{equation}
The charge susceptibility at $U=0$ adopts an RPA-like form
\begin{equation}
    \chi_c(q)\Big{|}_{U=0} = \frac{\chi_0}{1+Vf(\bq)\chi_0(q)}.
\end{equation}

\subsection{Results}

\begin{figure}
    \centering
    \includegraphics[width=0.99\linewidth]{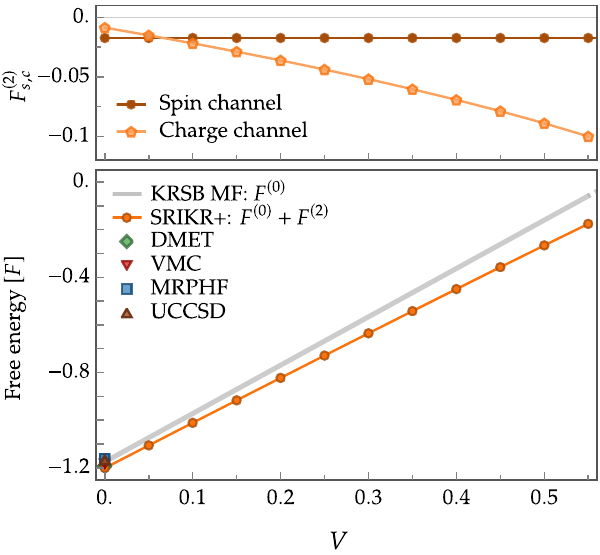}
    \caption{Fluctuation corrections to the free energy as a function of the long-range interaction $V$ for $U=2,t'=-0.2, n^{(0)}=1.0075$.}
    \label{fig:V}
\end{figure}

\autoref{fig:V} shows the free energy as a function of $V$ for $U=2$ near half-filling. The charge-channel fluctuation contribution increases with $V$, whereas the spin contribution is agnostic to $V$.

\section{Numerical evaluation of the CT contribution}\label{app:numerics}

There are three available methods to numerically evaluate the fluctuation corrections that have been derived in the main text, summarized by the following equations: 

(i) numerical evaluation of the Matsubara sum according to
\begin{align}\label{eq:app:OmegaCTMatsubara}
\begin{split}
    \Omega^{(2)}_{CT,s/c} &= \frac{a_{s/c}}{2}\frac{T}{N}\sum_{\bq}\sum_{\i\omega_n}\ln\det\mathcal{M}_{s/c}(q). 
\end{split}
\end{align}

(ii) numerical integration of the phase of the fluctuation determinant along the real axis according to
\begin{align}\label{eq:app:OmegaCT}
\begin{split}
\Omega^{(2)}_{CT,s/c} &= \frac{a_{s/c}}{2}\frac{1}{N}\sum_{\bq} \int_{0}^\infty \frac{d\omega}{\pi} \coth \left(\frac{\omega}{2T}\right)\arg\left(\det\mathcal{M}_{s/c}\right),
\end{split}
\end{align}
using that the functions $\ln\det\mathcal{M}_{s/c}$ are analytic on the complex plane except for the real axis and satisfy
\begin{align}
\ln\det\mathcal{M}_{s/c}(\bq,\omega + \i0^+) &= \left[\ln\det\mathcal{M}_{s/c}(\bq,\omega-\i0^+) \right]^{*}, \\
\ln\det\mathcal{M}_{s/c}(\bq,\omega + \i0^+) &= \ln\det\mathcal{M}_{s/c}(\bq,-\omega-\i0^+),
\end{align}
where the second identity holds because the determinants are even functions in $\omega_n$. Then the integration contour is moved to infinity and thereby wrapped around the real axis.

(iii) numerical integration along the real axis after partial integration according to
\begin{align}
\begin{split}
\Omega^{(2)}_{CT,s/c} &= \frac{a_{s/c}}{2}\frac{1}{N}\sum_{\bq} \int_{0}^\infty \frac{2 d\omega}{\pi} f(\omega) S_{s/c}(\omega),
\end{split}
\end{align}
where
\begin{align}
\begin{split}
S_{s/c}(\bq,\omega)&= 
\frac{-D_{s/c}'\partial_\omega D_{s/c}'' + D_{s/c}'' \partial_\omega D'}{|\det\mathcal{M}_{s/c}|^2}, \\
f(\omega)&= 
\frac{\omega}{2}+ T\log \big{|} 1-\e^{-\omega/T} \big{|},
\end{split}
\end{align}
and
\begin{align}
\begin{split}
D_{s/c}'&=\text{Re}\det\mathcal{M}_{s/c}(\bq,\omega+\i 0),\\
D_{s/c}''&=\text{Im}\det\mathcal{M}_{s/c}(\bq,\omega+\i 0). 
\end{split}
\end{align}
For each approach, there are numerical subtleties that we elaborate in the following. We are able to recover the same result with either of the methods, but there are substantial differences in the rate of convergence. The Matsubara summation is clearly favored over the frequency integrals. This is because the numerical error generated by a finite value of the convergence parameter $\eta > 0$, which is required for the integral method, is significant due to the overall small energy scale of the fluctuation corrections (in the order of $10^{-2}$). Therefore, we demand the numerical error to be $\Delta \Omega^{(2)}\ll 10 ^{-3}$. To achieve this precision, we require $\eta \lesssim 0.001$, which necessitates a very large number of fermionic momenta $N_\bk \gtrsim 1024$ because the imaginary part of the bare susceptibility $\Im \chi_0$ converges rather slowly for such small $\eta$. This problem does not occur for the Matsubara summation, since there is no convergence parameter required, and the bare susceptibility is a real-valued quantity in that case. Also, the Matsubara sum can efficiently be extrapolated to infinity, as we show below.

The sum over the bosonic momenta $\bq$ with a total number of $N_\bq$ also needs to be carried out numerically over the Brillouin zone (BZ). It can be reduced to $1/8$ of the BZ by exploiting the $c4v$ symmetry (4-fold rotation and mirror symmetry). There are no particular subtleties as in the case of the Matsubara summation, but $N_\bq$ needs to be large enough to encompass incommensurate fluctuation contributions, i.e, $N_\bq\gtrsim 8\times 8$ (compare \autoref{sec:incomm}).

\subsection{Numerical evaluation of the Matsubara sum}

In \eqref{eq:app:OmegaCTMatsubara}, we need to perform the summation $\sum_{\i\omega_n}=\sum_{n=-\infty}^\infty$ over the bosonic Matsubara frequencies $\omega_n = 2\pi n T$, where $T$ is the temperature. Since the fluctuation determinant is symmetric in $\omega_n$, this sum can be reduced to $n\geq 0$, whereby every point except $n=0$ is counted twice. The sum needs to be evaluated at finite temperature but can easily be extrapolated to $T\rightarrow 0$. We apply $T=0.01$, since the leading-order temperature-dependence is quadratic, i.e., the deviation from $T=0$ is only of the order $10^{-4}$.

\subsubsection{Spin sector}

After omitting the pre-factor $-p_0^2$ that drops out in the analytical continuation method (the sign is irrelevant and depends on the integration direction of the fields around the saddle point), the spin determinant is given by
\begin{equation}
\ln\det\mathcal{M}_s(q) = \ln\left[\chi_0(q)/\chi_s(q)\right] \sim a_{s,\bq}/\omega_n^2,    
\end{equation}
the asymptotic behavior of which guarantees convergence of the Matsubara sum. It may be evaluated approximately using a suitable cutoff $\omega_{n_c}=2\pi T n_c$. The proportionality factor is determined by $a_{s,\bq} =\lim_{\omega_{n}\rightarrow \infty} \ln\det\mathcal{M}_s(\bq,\omega_{n_c})\omega_{n_c}^2$. The asymptotic behavior can be summed up exactly as
\begin{widetext}
\begin{equation}
T\sum_{i\omega_n}\ln\det\mathcal{M}_s(\bq,\omega_n)
=
T\ln\det\mathcal{M}_s(\bq,0) 
+ 2T\sum_{n=1}^{n_c} \ln\det\mathcal{M}_s(\bq,\omega_n)
+T\frac{a_{s,\bq}}{4\pi^2 T^2} \psi^{(1)}(n_c+1)+O(1/n_c^3),
\end{equation}
\end{widetext}
where
\begin{equation}
\psi^{(1)}(n) = \sum_{x=n}^{\infty}\frac{1}{x^2}   
\end{equation}
is the Polygamma function and $n_c \rightarrow \infty$. In the numerical evaluation, we apply $n_c=3000$, $\omega_c = 2\pi T n_c$, whereby the error due to the cutoff is $\lesssim 10^{-5}$.

\subsubsection{Charge sector}

The charge sector is formally divergent in the ill-defined continuum limit, since 
\begin{equation}
\ln\det\mathcal{M}_c(q) \sim a_{c,\bq}\left(\omega_{\bq}^2+\omega_n^2 \right) \end{equation}
for $\omega_n\rightarrow \infty$. We may identify this behavior as a harmonic oscillator-like contribution of bosons of frequency $\omega_{\bq}$. The free energy of such a system is known to be
\begin{equation}
\Omega_{\text{free}}(\bq)=\omega_{\bq}/2+T\ln\left(1-e^{-\omega_{\bq}/T}\right).
\end{equation} 
We will separate the divergent part by fitting the above behavior beyond a cutoff $n_c$
\begin{widetext}
\begin{align}
T\sum_{i\omega_n}\ln\det\mathcal{M}_c(\bq,\omega_n)
&=
T\ln\frac{\det\mathcal{M}_c(\bq,0) }{a_{c,\bq} \omega_\bq^2}
+ 2T\sum_{n=1}^{\infty} \ln\frac{\det\mathcal{M}_c(\bq,\omega_n) }{a_{c,\bq}\left(\omega_\bq^2+\omega_n^2\right)}
+a_{c,\bq}\Omega_{\text{free}}(\bq).
\end{align}
\end{widetext}
In the numerical evaluation, we use a cutoff $n=n_c\approx 3000$ for the summation on Matsubara frequencies. The parameters $a_\bq,\omega_\bq$ are determined by the charge fluctuation determinant around the cutoff, where we define $\omega_1 = \omega_{n_c-1}, D_{1,\bq}=\det\mathcal{M}_c(\bq,\omega_1)$ and $\omega_2 = \omega_{n_c}, D_{2,\bq}=\det\mathcal{M}_c(\bq,\omega_2)$, yielding
\begin{align}
a_\bq &= \frac{D_{1,\bq}-D_{2,\bq}}{\omega_1^2-\omega_2^2},\\
\omega_\bq &= \sqrt{\frac{D_{2,\bq}\omega_1^2-D_{1,\bq}\omega_2^2}{D_{1,\bq}-D_{2,\bq}}},
\end{align}
such that $\ln\left[\det\mathcal{M}_c(\bq,\omega_{n_c})/a_\bq\left(\omega_\bq^2+\omega_{n_c}^2\right)\right]=0$ at the cutoff.

\subsection{Numerical evaluation of the frequency integral}

The integration along the real axis needs to be performed with care. First, the susceptibility $\chi_{0}(\bq,\omega+i0)$ entering the fluctuation matrix must be evaluated numerically, involving integration on $\bk$ over the Brillouin zone. This may be done by keeping a finite imaginary part $\eta$ of the frequency, $\omega \rightarrow \omega + i\eta$ to ensure convergence of the numerical integration. The parameters $\eta$, the step size $d\omega$ of the discretized frequency and the temperature $T$ need to be chosen judiciously. We choose both, $d\omega \ll \eta \ll T$ and sufficiently high number of points $N_\bq,N_\bk$ in momentum space. 

\subsubsection{Phase integration}

The CT fluctuation contribution to the partition function is determined by integrating the phase of the fluctuation determinant weighted by the Bose function 
\begin{align}\label{eq:OmegaCTappendix}
\begin{split}
\Omega^{(2)}_{CT,s/c} &= a_{s/c}\frac{1}{N}\sum_{\bq} \int_{0}^\infty \frac{d\omega}{2\pi} \coth \left(\frac{\omega}{2T}\right)\arg\left(\det\mathcal{M}_{s/c}\right).\\
\end{split}
\end{align}
where $\arg\left(\det\mathcal{M}_{s/c}\right)$ is the phase of $\det\mathcal{M}_{s/c}$.
The phase of the spin determinant is given by the usual expression 
\begin{align}
\arg\left(\det\mathcal{M}_{s}\right) =\arctan\left(D_{s}''/D_{s}'\right)
\end{align}
where we defined the real and imaginary parts of the determinant $\det\mathcal{M}_{s/c}=D_{s,c}'+iD_{s,c}''$. 
The charge determinant features roots on the real axis such that the continuity of the phase needs to be recovered by the correct addition of $\pi$. For $T=0$, this is achieved by  
\begin{align}
\arg\left(\det\mathcal{M}_s\right) &=\,\,\, \arctan\left(D_s''/D_s'\right),\\
\arg\left(\det\mathcal{M}_c\right)&=
\begin{cases}
\arctan\left(D_c''/D_c'\right) \qquad\,\,\, \text{if } D_c'>0,D_c''>0 \\  
\arctan\left(D_c''/D_c'\right)+\pi \quad \text{else}. \nonumber
\end{cases}
\end{align}

\subsubsection{Partial integration}
Alternatively, we can perform a partial integration, which avoids the issue with the continuity of the phase:
\begin{align}
\Omega^{(2)}_{CT,s/c} &= a_{s/c}\frac{1}{N}\sum_{\bq} \int_{0}^\infty \frac{d\omega}{\pi} f(\omega) S_{s/c}(\omega),\label{eq:OmegaCTss}
\end{align}
where we define the functions
\begin{align}
\begin{split}
S_{s/c}(\bq,\omega)&= -\text{Im} \frac{\partial_\omega \det\mathcal{M}_{s/c}(\bq,\omega+i0^+)}{\det\mathcal{M}_{s/c}}\\
&= \frac{-D_{s/c}'\partial_\omega D_{s/c}'' + D_{s/c}'' \partial_\omega D'}{|\det\mathcal{M}_{s/c}|^2} f(\omega), \\
f(\omega)&= \frac{\omega}{2}+ T\log \big{|} 1-\e^{-\omega/T} \big{|}.
\end{split}
\end{align}
 However, the integration converges slower than the phase integration due to the numerical derivative that needs to be evaluated. Due to $\eta\rightarrow 0$ remaining finite within the numerics, \eqref{eq:OmegaCTss} features a logarithmic divergence for $\omega\rightarrow\infty$ in the charge sector. Convergence can be recovered by introducing a cutoff $\omega_c$ of the integral that is chosen such that the first order of its $\eta$-dependence vanishes, i.e., $\Omega^{(2)}_{CT,c} = a + b \eta^3 + \mathcal{O}(\eta^5)$.

\subsection{Comparison of HFC and CT contribution}

\begin{figure}
    \centering
    \includegraphics[width=0.99\linewidth]{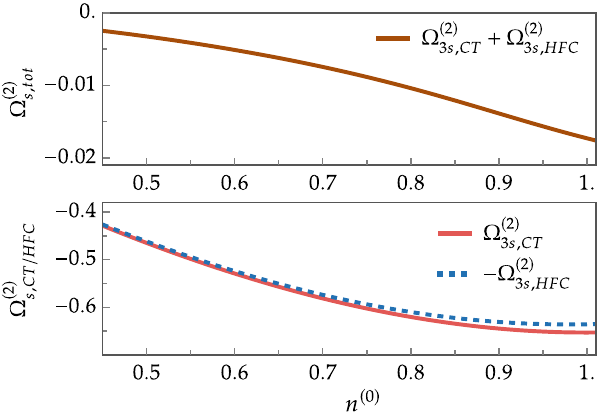}
    \caption{Spin fluctuation contributions to the grand potential as a function of the MF density $n^{(0)}$ for $U=2,t'=+0.2,T=0.01$. The bottom panel shows the CT part and minus the HFC separately.}
    \label{fig:OmegaCTHFCs}
\end{figure}

\begin{figure}
    \centering
    \includegraphics[width=0.99\linewidth]{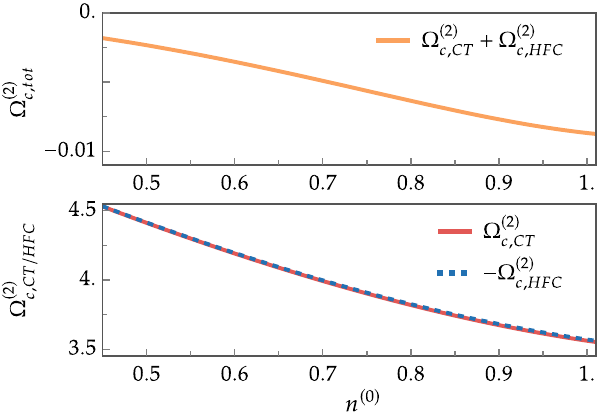}
    \caption{Charge fluctuation contributions to the grand potential as a function of the MF density $n^{(0)}$ for $U=2,t'=+0.2,T=0.01$. The bottom panel shows the CT part and minus the HFC separately.}
    \label{fig:OmegaCTHFCc}
\end{figure}

In \autoref{fig:OmegaCTHFCs} and \autoref{fig:OmegaCTHFCc}, we show the results for the HFC and CT contribution separately in the lower panels, as well as their sum, that is, the total result, in the upper panels.
In case of spin, the total fluctuation contribution is reduced to less than $3\%$ of the CT result with the addition of the HFC. For the charge sector, the situation is even more extreme, since the CT part is of the order of $\overline{\alpha}$ (e.g. for $n=1,U=0,t'=0$, $\overline{\alpha}=64/\pi^2$) and the correct non-interacting limit is only recovered with the addition of the HFC, compare \autoref{sec:app:U0}. Here, with the addition of HFC, the total charge fluctuation result is reduced to less than $0.25\%$ of the CT value. These results illustrate the huge errors caused by the CT approximation that, by all means, need to be corrected with the addition of HFCs to recover an error-free functional integral representation.

\section{Tables of numerical results}

Here, we provide the numerical values of the SRIKR+ (mean field plus fluctuation) results that are shown within \autoref{fig:FEU=2} and \autoref{fig:n}. The fluctuation calculation is done for the grand canonical ensemble ,i.e., we use the chemical potential $\mu_0$ as an input parameter and calculate the density $n_{tot}=n^{(0)}+n^{(2)}_s+n^{(2)}_c$ afterwards.
The data shown in \autoref{tab:data} and \autoref{tab:data2} is extrapolated to fractions of the total filling with negligible error. 
The total free energy is given by the sum of the mean field and fluctuation contribution $F_{\text{tot}}=F^{(0)}+F^{(2)}_s+F^{(2)}_c$.

Our method requires a finite temperature because of the Matsubara summation. We applied $T=0.01$ as zero temperature limit. This causes a difference of less than $10^{-4}$ wrt.~to the free energy, since we find the temperature coefficient $\gamma \lesssim 1$. The numerical difference of the free energy at $n=1$ calculated for $t=\pm0.2$ is of the order of only $10^{-6}$ (the result must be identical due to particle-hole symmetry). This underlines the effectiveness of our numerical calculations. 

\begin{table} 
    \centering
    \caption{SRIKR+ results for $U=2, t'=-0.2$} 
    \begin{tabular}{c@{\hspace{20pt}} c@{\hspace{20pt}} c@{\hspace{20pt}} c@{\hspace{20pt}} c} 
        \toprule
        $n_{tot.}$  & $F_{tot.}$  & $n^{(0)}$  & $F^{(0)}$  & $\mu_0$  \\
        0.5  & -1.0687  & 0.4872  & -1.0434  & -1.1277  \\
        0.6  & -1.1631  & 0.5865  & -1.1358  & -0.7625  \\
        0.7  & -1.2221  & 0.6841  & -1.1936  & -0.4224  \\
        0.8  & -1.2487  & 0.7851  & -1.2199  & -0.1148  \\
        0.9  & -1.2442  & 0.9055  & -1.2145  & 0.2150   \\
        1.0  & -1.2051  & 1.0109  & -1.1733  & 0.5745   \\
        \hline\hline 
    \end{tabular}
    \label{tab:data}
\end{table}

\begin{table} 
    \centering
    \caption{SRIKR+ results for $U=2, t'=+0.2$} 
    \begin{tabular}{c@{\hspace{20pt}} c@{\hspace{20pt}} c@{\hspace{20pt}} c@{\hspace{20pt}} c} 
        \toprule
        $n_{tot.}$  & $F_{tot.}$  & $n^{(0)}$  & $F^{(0)}$  & $\mu_0$  \\
        0.5  & -1.3666  & 0.4954  & -1.3567  & -0.9670  \\
        0.6  & -1.4341  & 0.5939  & -1.4232  & -0.3934  \\
        0.7  & -1.4469  & 0.6920  & -1.4359  & 0.1273   \\
        0.8  & -1.4101  & 0.7898  & -1.3999  & 0.5999   \\
        0.9  & -1.3282  & 0.8883  & -1.3193  & 1.0307   \\
        1.0  & -1.2051  & 0.9893  & -1.1948  & 1.4264   \\
        \hline\hline 
    \end{tabular}
    \label{tab:data2}
\end{table}

%%%%%%%%%%%%%%%%%%%%%%%%%%%%%%%%%
%          Bibliography         %
%%%%%%%%%%%%%%%%%%%%%%%%%%%%%%%%%

\bibliography{bibliography}

\end{document}